\documentclass[12pt, preprint]{aastex}
\usepackage{times}
\newcommand{\etal}{{et al}\/.}

\begin{document}
\slugcomment{ApJ accepted May 25 2004}
\shorttitle{X-ray hotspots}
\shortauthors{M.J.\ Hardcastle \etal}
\title{The origins of X-ray emission from the hotspots of FRII radio sources}
\author{M.J.\ Hardcastle}
\affil{Department of Physics, University of Bristol, Tyndall Avenue,
Bristol BS8 1TL, UK}
\author{D.E.\ Harris}
\affil{Harvard-Smithsonian Center for Astrophysics, 60 Garden Street, Cambridge, MA~02138, USA}
\and
\author{D.M. Worrall and M. Birkinshaw}
\affil{Department of Physics, University of Bristol, Tyndall Avenue,
Bristol BS8 1TL, UK}

\begin{abstract}
We use new and archival {\it Chandra} data to investigate the X-ray
emission from a large sample of compact hotspots of FRII radio
galaxies and quasars from the 3C catalogue. We find that only the most
luminous hotspots tend to be in good agreement with the predictions of
a synchrotron self-Compton model with equipartition magnetic fields.
At low hotspot luminosities inverse-Compton predictions are routinely
exceeded by several orders of magnitude, but this is never seen in
more luminous hotspots. We argue that an additional synchrotron
component of the X-ray emission is present in low-luminosity hotspots,
and that the hotspot luminosity controls the ability of a given
hotspot to produce synchrotron X-rays, probably by determining the
high-energy cutoff of the electron energy spectrum. It remains
plausible that all hotspots are close to the equipartition condition.
\end{abstract}

\maketitle
\section{Introduction}
\label{intro}
The hotspots of powerful FRII radio sources, as observed in the radio,
have long been believed to be the observable consequence of a strong
terminal shock at the end of the relativistic jets that feed the radio
lobes. This picture has to be modified somewhat in the light of the
fact that a large number of lobes contain more than one hotspot when
observed at high resolution \citep{l82,lbdh97,hapr97}. Conventionally
the most compact feature, which is universally the one that lies at
the end of the jet when the jet termination can be observed, is called
the `primary' hotspot, while the others are known as `secondary'
hotspots. Since standard shock acceleration and energy loss models are
often good fits to the radio-to-optical spectra of both primary and
secondary hotspots \citep[e.g.,][]{mrhy89}, it seems likely that in
many cases the secondary hotspots are also being powered by bulk
kinetic energy from the jet.

For some time it has been clear that it is not possible to explain the
X-ray emission from hotspots with a single model. Some objects show
hotspot X-ray emission with a spectrum consistent with the predictions
of a synchrotron self-Compton (SSC) model, in which the
synchrotron-emitting electrons inverse-Compton (IC) scatter
synchrotron photons into the X-ray band; in these objects there is
good agreement between the observed flux density and the predictions
of an SSC model with a magnetic field close to the value expected for
equipartition of energy between the magnetic field and the radiating
electrons, and so their X-ray emission is attributed to the SSC
process (e.g. \citealt*{hcp94}; \citealt{hbch02}, hereafter H02). On
the other hand, objects such as 3C\,390.3 \citep*{hll98} and Pictor~A
\citep{rm87} show X-ray emission that is clearly much stronger than
the SSC model would predict if the magnetic field had the
equipartition value, together with a spectrum that is steeper than the
low-frequency radio spectrum, and hence too steep to be
inverse-Compton (\citealt*{wys01}; H02). In some, but not all, of
these cases, a simple synchrotron spectrum (by which we mean a single
power law or a broken, steepening power law in frequency) is a good
fit to the radio, optical and X-ray data points. In addition, there
are several sources (the best example being 3C\,351: H02) where the
X-ray structure is clearly different from that seen in the radio maps,
which is impossible in a simple SSC model with a homogeneous magnetic
field and electron distribution. In a synchrotron model for some or
all of the X-rays, differences in the spatial structure are to be
expected, since the synchrotron loss timescale for X-ray emitting
electrons (tens of years in a typical equipartition magnetic field) is
orders of magnitude less than that for radio-emitting electrons ($\ga
10^5$ years); in fact, in a non-steady-state situation (as expected
from numerical simulations, e.g. \citealt*{tjr01}) time-varying
differences in both spatial and spectral distributions of the radio
and X-ray emitting electrons are more or less required by the physics.

Until now, however, it has not been clear {\it why} some hotspots'
X-ray emission is adequately modeled by the SSC process with an
equipartition field, while others require an additional component or a
lower than equipartion field strength. Suggested explanations have
involved (1) a lower magnetic field strength in the X-ray bright
hotspots, which both increases IC emission and increases the loss
lifetime of X-ray synchrotron-emitting electrons \citep{bbcp01}; or
(2) the effects of differential relativistic beaming, due to
decelerating bulk motions in the hotspots, on the synchrotron and IC
spectra, in particular the fact that fast-moving parts of the flow see
the slow-moving downstream flow Doppler-boosted \citep{gk03}, which
helps to account for the fact that many of the early detections of
X-ray bright hotspots involved broad-line radio galaxies or quasars
(H02), which in unified models should lie relatively close to the line
of sight. The small number of published detections of X-ray hotspots
has made it difficult to arrive at a definitive answer.

Since the appearance of earlier work aimed explicitly at detecting
hotspots \citep{hnpb00,hbw01a,hbch02,bbcs02} a number of hotspots have
been detected in {\it Chandra} observations of FRII radio galaxies and
quasars made for other purposes \citep[e.g.,][]{ddh03,cf03}. This has
motivated us to collate all existing data on the hotspots of 3C radio
sources from the {\it Chandra} archive and analyse them in a
systematic way, with the aim of determining trends and testing models.
In this paper we report our results.

Throughout the paper we use a cosmology in which $H_0 = 70$ km
s$^{-1}$ Mpc$^{-1}$, $\Omega_{\rm m} = 0.3$ and $\Omega_\Lambda =
0.7$. Spectral indices $\alpha$ are the energy indices and are defined
in the sense $S_{\nu} \propto \nu^{-\alpha}$.

\section{Data and analysis}

We searched the public {\it Chandra} archives for all observations of
FRII radio sources in the 3C catalogue made with the CCD Imaging
Spectrometer (ACIS), supplementing them with a few observations that
we have access to and are not yet public. To make our sample as large
as possible, we included any 3C FRII source listed in the archives;
this means that we made use of several sources that are not in the
better-defined 3CR \citep{sdma85} or 3CRR \citep{lrl83} samples. We
restricted ourselves to the 3C sources simply because they almost all
have good radio observations available in the public NRAO Very Large
Array (VLA) archives. We excluded compact steep-spectrum sources whose
hotspots would not be resolved from the AGN or lobe emission with {\it
Chandra}. These selection criteria gave us a sample of 36 sources
(Table \ref{obs}). We next obtained electronic radio maps for all the
sources in our sample. Where we did not already have access to a good,
high-resolution radio map, or to appropriate published radio flux
densities, we retrieved the best available data from the VLA archive.
In selecting the observations we preferred data at 5 and 8 GHz in the
A configuration (the largest configuration of the VLA), which gives
good sensitivity, good separation of lobes and hotspots due to their
spectral index differences, and an angular resolution comparable to
that of {\it Chandra}. Properties of the radio data used and maps made
are given in Table \ref{radioobs}.

We then used the radio maps as a guide to search for emission from
hotspots. We defined a hotspot less strictly than some other workers
\citep[e.g.,][]{bhlb94} both in order to make the comparison between
radio and X-ray simpler and because our aim is to include the
structure apparently associated with the jet termination whenever
possible. In practice, we considered any relatively compact, isolated
radio feature that was significantly brighter than its surroundings to
be a hotspot \citep[cf.][]{lbdh97}. However, we excluded any emission
that we considered to be associated with a jet; any compact X-ray
feature positionally coincident with, or closer to the nucleus than, a
linear radio feature that met the definition of \citet{bp84}, was
considered to be a `jet knot' rather than a hotspot. By doing this we
hoped to select only features associated with the termination of the
jet, and to avoid effects thought to be due to highly relativistic
bulk motions, as seen in the X-ray jets of some quasars
\citep[e.g.,][]{tmsu00}. Later in the paper (\S\ref{jk}) we shall
return to the question of whether hotspots tell us anything about the
emission from jet knots or jets in general.

Using the radio data as a guide, we were able to identify a number of
previously unreported X-ray counterparts to hotspots, in the 10 FRIIs
3C\,6.1, 3C\,47, 3C\,109, 3C\,173.1, 3C\,228, 3C\,321, 3C\,324,
3C\,334, 3C\,403 and 3C\,452. The results on 3C\,403 and 3C\,228, 334
will be reported in more detail elsewhere (respectively in R.\ Kraft
\etal , and D.M.\ Worrall \etal , in preparation); images of the newly
detected X-ray hotspots for the other sources are presented in
Appendix A. In almost all cases there were a few tens of total counts
in the hotspots in the 0.5--5 keV range. Since this is too few to fit
spectra, we used the standard {\it Chandra} analysis software CIAO to
generate redistribution matrix (RMF) and ancilliary response (ARF)
files appropriate for the hotspots, using the {\sc psextract} tool,
corrected for the time-dependent excess ACIS absorption using {\sc
apply\_acisabs}, and then used the model-fitting software XSPEC to
determine the normalization of a power law, with $\alpha = 0.5$ and
Galactic absorption, that reproduced the observed net count rate. We
chose this power-law index because our aim was to test the validity of
the SSC model, which `predicts' $\alpha = 0.5$, on the basis of the
assumption that the low-energy electron energy index has the value 2.0
associated with particle acceleration at a non-relativistic strong
shock (as appears to be the case in some, though not all, well-studied
hotspots: \citealt*{myr97}). However, the choice of $\alpha$ makes
relatively little difference to the normalization of the power law,
and thus to the inferred 1-keV flux density in the observer's frame;
using $\alpha = 1$ would increase the inferred flux density by between
10 and 20\%. The 1-keV flux densities assuming $\alpha=0.5$ for each
source are given in Table \ref{fluxes}.

For the 10 sources 3C\,9, 3C\,184, 3C\,200, 3C\,212, 3C\,215, 3C\,219,
3C\,220.1, 3C\,401, 3C\,427.1 and 3C\,438 we found no X-ray emission
associated with the hotspots, as previously reported in some cases.
Where there was a compact (arcsec or sub-arcsec) radio hotspot present
we determined $3\sigma$ upper limits on the corresponding X-ray flux
density in a 10-pixel detection cell based on Poisson statistics and
the local background count rate. Some of these sources (e.g. 3C\,401
and 3C\,438, \citealt{hapr97}) exhibit no compact hotspots in
the radio, and we elected not to determine an upper limit on
their emission, as the selection of an appropriate X-ray region is
difficult. For the same reason, we did not determine an upper
limit for the barely resolved source 3C\,184, which shows X-ray
emission coincident with the radio lobe but not particularly the
hotspot \citep{bwhb04}.

We re-examined previously reported hotspot detections in the 16
sources 3C\,123 \citep{hbw01a}, 3C\,179 \citep{smtu02}, 3C\,207
\citep{bbcs02}, 3C\,254 \citep{ddh03}, 3C\,263 (H02), 3C\,265
\citep{bbcs03}, 3C\,275.1 \citep{cf03}, 3C\,280 \citep{ddh03}, 3C\,281
\citep{cf03}, 3C\,294 \citep{fsce03}, 3C\,295 \citep{hnpb00}, 3C\,303
\citep{kegt03}, 3C\,330 (H02), 3C\,351 (\citealt{bbcp01}; H02),
3C\,390.3 \citep{hll98}, and Cygnus A (3C\,405: \citealt{hcp94}). In
all but one case we confirmed the existence of one or more compact
X-ray features associated with the radio hotspots (the exception is
3C\,281, where the previously reported X-ray emission appears to be
diffuse and associated with the lobe, and is most likely due to IC
scattering of cosmic microwave background (CMB) photons by the lobe
rather than SSC from the hotspots). Where a 1-keV flux density had
been previously determined from spectral fitting, we make use of that
in Table \ref{fluxes}. Otherwise, we adopted the same procedure as for
the newly detected sources described above. Finally, for all the newly
detected and known sources, we determined upper limits, again as
described above, for any compact hotspots that were not detected (e.g.
in the lobe on the opposite side of the nucleus to the known hotspot).
In the process of doing this we found one additional hotspot, 3C\,123
W, that was formally significantly detected; although the situation in
this source is confused by the presence of strong, unrelaxed cluster
emission \citep{hbw01a} we added it to the sample as a detection, for
consistency with the other sources. Because of its intrinsic interest
and the extreme nature of its X-ray hotspot, we added Pictor A
\citep*{wys01} to the sample (it is not in 3C because of its low
declination, but meets the other selection criteria). The overall
final sample thus contains 37 sources (Table \ref{obs}) with 65 X-ray
hotspot flux densities or upper limits. All the fluxes and upper
limits are tabulated in Table \ref{fluxes}.

Finally, for the sources with detected X-ray hotspots, we used the
{\it Hubble Space Telescope} ({\it HST}) archive to search for optical
counterparts. Optical emission is important because it constrains the
spectrum between radio and X-ray; early work on SSC hotspots was
supported by the observation that a one-zone synchrotron model could
not be fitted through the radio and X-ray data points because of the
optical constraints \citep[e.g.,][]{hcp94}. We identified two new
candidate optical hotspot counterparts (in 3C\,228 and 3C\,275.1) and
measured flux densities or upper limits for a number of other sources,
using the IRAF package {\sc synphot} to calculate the conversion
factor between observed counts and flux density. Optical flux
densities and frequencies are tabulated in Table \ref{hst}. Sources
where there were no archival {\it HST} observations, where the hotspot
did not lie on the WFPC-2 CCDs, or where observational constraints
such as a nearby bright star or cosmic ray contamination prevented us
from obtaining a flux density, are not tabulated. We also tabulate a
number of flux densities, largely based on ground-based observations,
taken from other papers, either in the literature or in preparation.

\section{Modeling and results}
\label{model}

The large number of detected X-ray hotspots is interesting in itself,
given that calculations based on SSC emission at equipartition
suggested that only the few brightest hotspots would be detected with
{\it Chandra} \citep[e.g.,][]{h01}. In order to assess quantitatively
the extent to which the new detections conflict with an SSC model, we
decided to fit a simple SSC model to all the hotspots and determine
the ratio between the observed and predicted flux densities. We
carried out this calculation using the code of \citet*{hbw98}; a brief
sketch of the operation of this code is given in Appendix B. The code
assumes a spherically symmetric, homogenous hotspot with an electron
energy spectrum that can be described as a power law or broken power
law. To determine the radius of the hotspot, we therefore fitted
models consisting of a homogeneous sphere convolved with the restoring
beam to the highest-resolution radio data available, in the manner
described by H02. Where multi-frequency radio data were available for
the hotspot, which was only true in the best-studied cases, we used
them to fit a two-component power-law model with an energy spectral
break corresponding to $\Delta\alpha = 0.5$ \citep{hm87}; otherwise we
assumed a single power law with $\alpha=0.5$ extending from the radio
into the mm-wave regime. Unless good low-frequency radio constraints
were available, we assumed that the minimum Lorentz factor of the
electrons, $\gamma_{\rm min}$, was 1000; $\gamma_{\rm max}$ was chosen
to ensure that there was no spectral cutoff before the mm-wave region.
These choices reflect what has been found in the best-studied
hotspots, but are clearly no substitute for good, multi-frequency
observations, particularly at high frequencies. However, we estimate
that these choices make a difference at the level of at most $\sim
10$--20\% (except in the rare cases where a $\sim 10$ GHz spectral
cutoff is present). The results are particularly insensitive to the
choice of $\gamma_{\rm min}$, since decreasing this has two effects
which act in opposite directions: more high-energy photons are
scattered by the large additional population of low-energy electrons,
but the overall electron energy power law normalization is reduced to
maintain equipartition. Detailed spatial modeling, where
high-resolution observations make it possible, also changes the
results of SSC calculations at the 10--20\% level (H02), so that
overall the calculated value should be a good estimate of the true
inverse-Compton prediction. The equipartition flux density prediction
(taking into account both SSC and IC scattering of CMB photons) and
the ratio $R$ between the observed and predicted flux densities is
tabulated in Table \ref{fluxes} for each hotspot. Note that in almost
all cases the flux due to the SSC process dominates over that due to
IC scattering of the CMB (assuming no relativistic beaming) by an
order of magnitude or more. For simplicity we shall often refer to the
calculated fluxes as SSC fluxes in what follows.

The tabulated values of $R$ are calculated assuming that there are no
protons, so that equipartition is between the radiating electrons and
magnetic fields only, and that the filling factor of the hotspots is
unity. If we were to include an energetically dominant population of
protons in equipartition in our model, it would reduce the predicted
inverse-Compton emission (since the number of electrons decreases) and
so increase the ratio $R$, possibly by a large factor. If the ratio of
proton to electron energy densities were in the ratio of their rest
masses, $R$ values would increase by about a factor 30--70. Even if
the proton to electron energy densities were of the order of their
number ratios as observed in cosmic rays at the Earth (an
energy-dependent factor of $\sim 50$--$100$, \citealt{l92}) we would
expect $R$ to increase by a factor $\sim 5$--$10$. A population of protons
with the {\it same} total energy as the electrons has a less dramatic
effect, increasing $R$ by only about 40\%. In any case, it is clear
that introducing protons cannot solve the problem of high-$R$
hotspots. The predicted SSC inverse-Compton emission can be increased,
and $R$ can be decreased, if we have overestimated the volume or the
filling factor, though (depending on the space-filling fluid) the
actual results of a low filling factor can be very geometry-dependent
-- if the electrons are confined to thin sheets, so that the
probability of scattering is comparatively low, then the effects of
low filling factor can be less than expected. Roughly (see Appendix
B), to reduce $R$ by a factor of 1000, and so to make the most extreme
observed hotspots consistent with being inverse-Compton emission at
equipartition, we would need to reduce the volume or the filling
factor by a factor $\sim 10^{12}$, and this neglects geometrical
effects. Such low filling factors are clearly implausible.

\begin{figure*}
\begin{center}
\plottwo{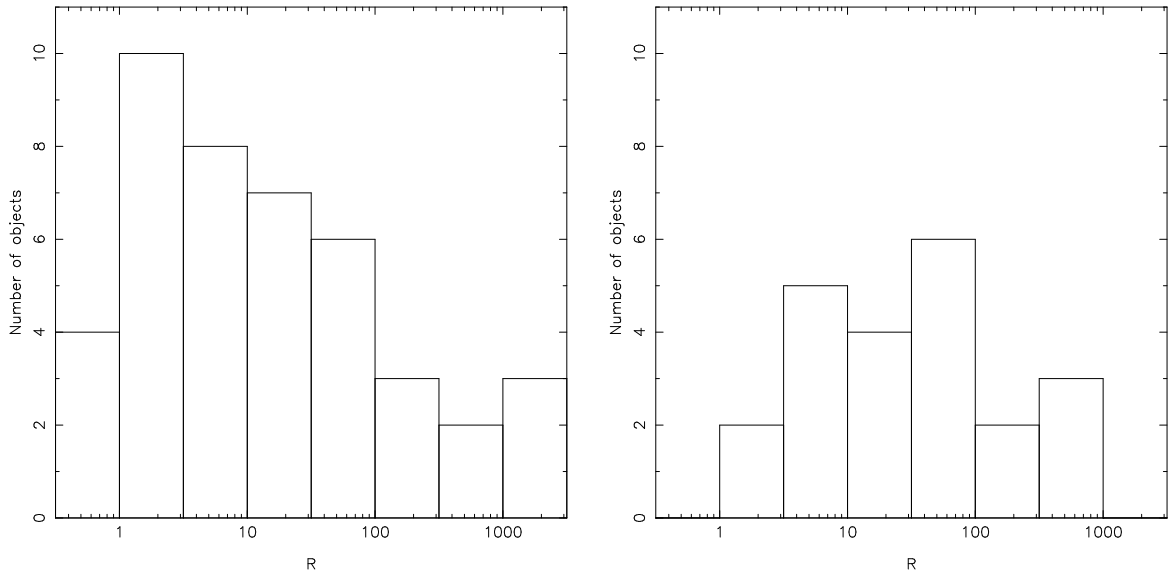}{f1b.eps}
\end{center}
\caption{Distribution of the ratio $R$ of observed X-ray flux to IC
  prediction for the {\it Chandra} sample. Left: distribution for
  X-ray detected hotspots. Right: distribution for compact hotspots
  with no X-ray detection (sources can move to the left).}
\label{histo}
\end{figure*}

Fig.\ \ref{histo} shows the distribution of the ratio $R$ and upper
limits on $R$ for the sample. Two points are immediately obvious:
firstly, the detected X-ray flux density lies significantly above the
IC prediction in most sources; secondly, there are few detected
sources with $R<1$, and there is a clear change in the distribution of
sources at around $R = 1$. If the upper limits on X-ray flux for the
non-detected sources all lie a long way above the true values, then
this could change, but if it does not, the special status of $R=1$
implies that few sources have {\it less} X-ray emission than would be
expected on the equipartition SSC/IC model. This could suggest either
that all sources have SSC/IC X-ray emission at a level consistent with
the equipartition prediction, together with some additional source of
X-ray emission, or that there are departures from equipartition ---
some quite large --- but that these are always in the sense that
$B<B_{\rm eq}$. The fact that we do not see many sources with $R\ll 1$
suggests that there are few or no hotspots with $B\gg B_{\rm eq}$,
though without detections of all the hotspots we cannot be more
definite.

If the model we have used to predict the level of inverse-Compton
emission is incorrect, then the special status of $R=1$ would have to
be a coincidence. This, as we have argued before (e.g.\ H02) gives us
a reason to disfavor models with an energetically dominant proton
population, or with consistently very low filling factor. Thus, for
example, if the proton-to-electron energy density ratio were $\sim
100$, we would expect a source in equipartition, and emitting in X-rays
only via the SSC process, to have $R \approx 0.1$--$0.2$ given our model
(since we would be {\it overpredicting} the inverse-Compton emission);
we would not see a special status for $R = 1$ unless other
parameters (such as filling factor) conspired systematically to
increase the value of $R$, which is inherently improbable. We
emphasise that, given the small numbers and the width of the
distribution of $R$ values, this
does not rule out moderate filling factors or a proton population
within a factor of $\sim 10$ of the energy density in the other
components.

\begin{figure*}
\begin{center}
\plotone{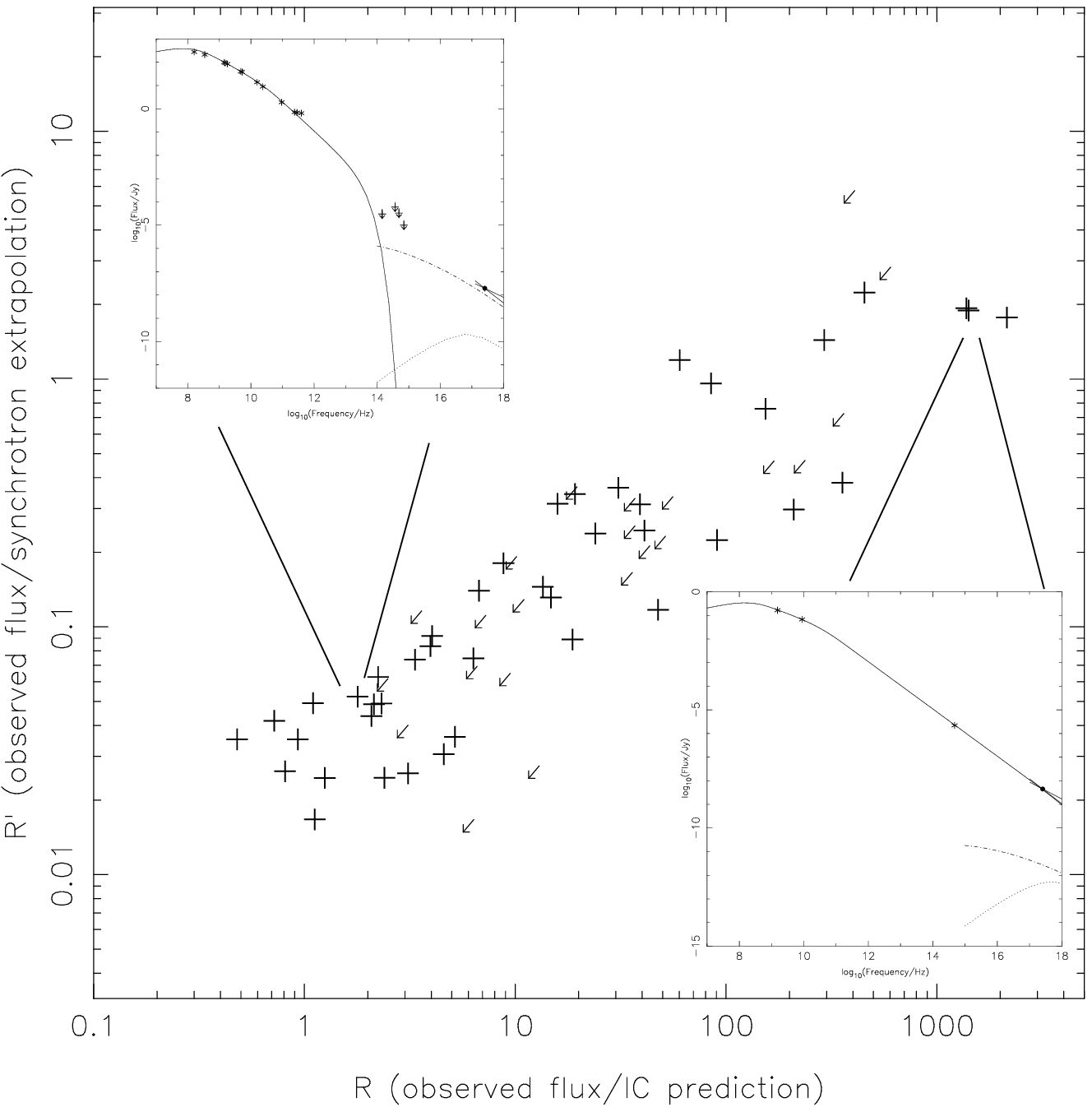}
\end{center}
\caption{$R'$ is plotted against $R$, where $R$ is the ratio between
the observed X-ray flux density and the prediction of an
inverse-Compton model at equipartition, and $R'$ is defined as the
ratio of the observed flux to the extrapolation of the radio flux
density assuming $\alpha = 1.0$, i.e. to the amount of X-rays that
could (conservatively) have been produced by synchrotron emission with
a straight spectrum. Diagonal arrows show upper limits from
non-detected X-ray sources. Insets show the broad-band SEDs of two
extreme sources on the plot: top left, Cygnus A hotspot A; bottom
right, 3C\,390.3, N hotspot. The data points are from the literature
or from maps available to us, the solid line represents the
best-fitting synchrotron model, and the dot-dashed and dotted lines
represent the equipartition synchrotron self-Compton and CMB
inverse-Compton models, respectively.}
\label{rrp}
\end{figure*}
\begin{figure*}
\begin{center}
\plotone{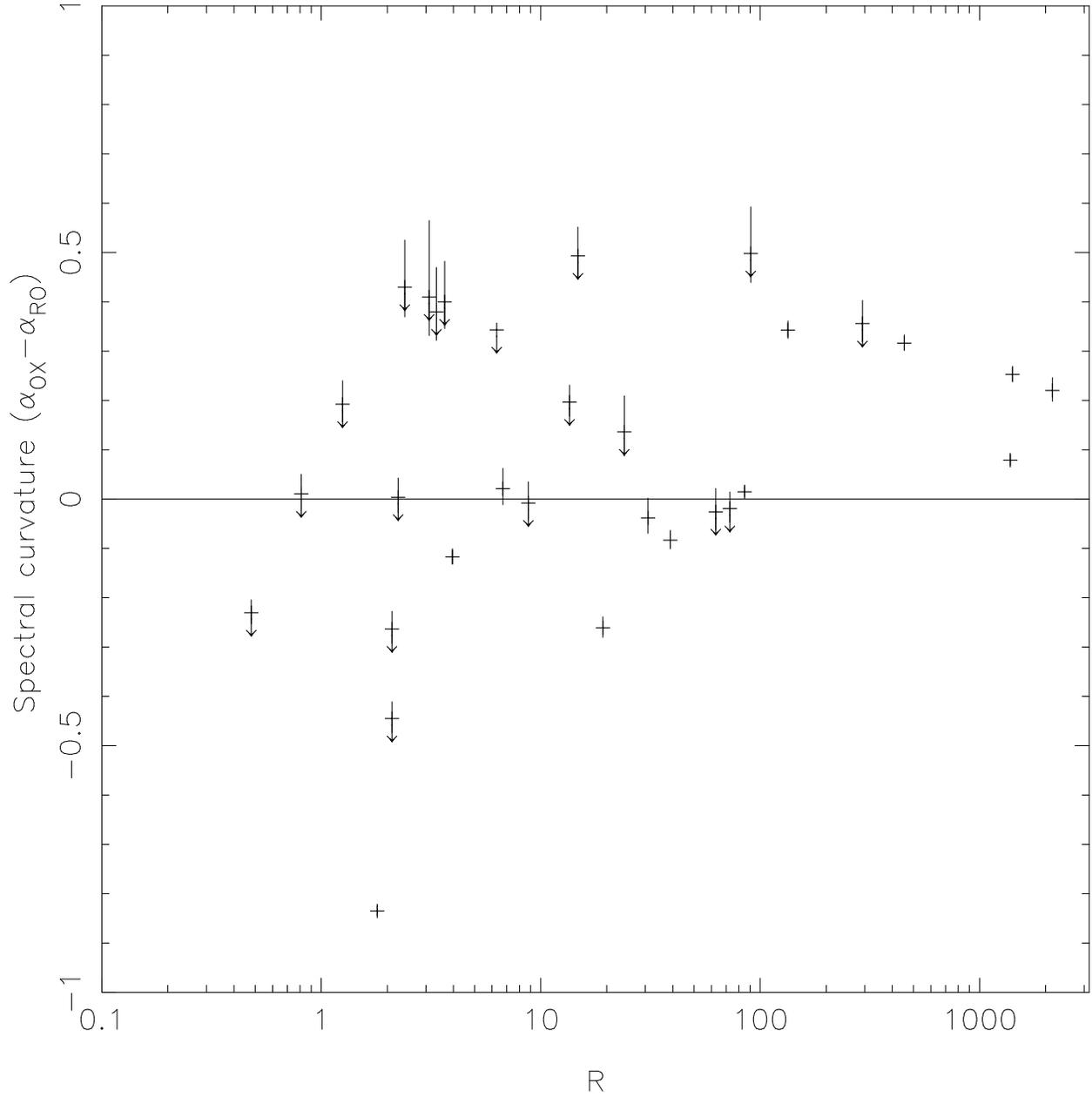}
\end{center}
\caption{Spectral index difference (curvature indicator) against $R$
  for sources with optical hotspots or upper limits on optical flux.
  The error bars show the statistical errors on X-ray flux density
  only, as these are the dominant errors.}
\label{optical}
\end{figure*}

What emission mechanisms are possible for the detected hotspots? We
began by calculating another parameter, $R'$, the ratio between the
observed X-ray flux density and the flux predicted from a simple
power-law extrapolation (with $\alpha_{\rm RX} = 1.0$) from the radio
data. Hotspots with both $R\gg1$ and $R'\gg1$ would represent a
problem for both synchrotron and inverse-Compton models. However, we
find that $R'$ is almost always less than 1, so there are no sources
whose X-ray flux is impossible to explain with a synchrotron model in
this sense. The plot of $R'$ against $R$ (Fig. \ref{rrp}) shows that
even the most extreme X-ray hotspots, in terms of $R$, can readily be
accounted for with a synchrotron model. There is a smooth distribution
in parameter space, with no obvious bimodality, between sources with
$R'=1$, $R\gg1$ (where a synchrotron model is natural) and $R=1$,
$R'\ll 1$ (where an inverse-Compton model has tended to be adopted in
earlier work). Insets in Fig. \ref{rrp} show the very different SEDs
of sources at the extreme ends of the distribution, and also
illustrate the importance of optical constraints in determining the
X-ray emission mechanism. To investigate this further, we used the
available optical data or upper limits for X-ray detected hotspots to
constrain their spectral shape. We calculated the quantities
$\alpha_{\rm RO}$ and $\alpha_{\rm OX}$, the two-point radio to
optical and optical to X-ray spectral indices, for all the X-ray
detected hotspots with optical flux densities or upper limits. A
hotspot in which the optical to X-ray spectrum is flatter than the
radio to optical spectrum ($\alpha_{\rm OX} < \alpha_{\rm RO}$) cannot
be described by a simple one-zone synchrotron model in which the
spectrum steepens with increasing frequency. The difference between
the two spectral indices is plotted in Fig.\ \ref{optical} as a
function of $R$. It can be seen that the more extreme hotspots (large
$R$ values) all have $\alpha_{\rm OX} - \alpha_{\rm RO}>0$, and so are
consistent with a synchrotron model. A non-synchrotron model is
required by the optical data only for a few low-$R$ objects, where
inverse-Compton emission is the accepted and most plausible X-ray
mechanism. The existing data do not rule out a model in which
synchrotron X-ray emission is important in a significant number of our
target objects, although this is far from conclusive given the large
number of optical non-detections.

\begin{figure*}
\begin{center}
\plotone{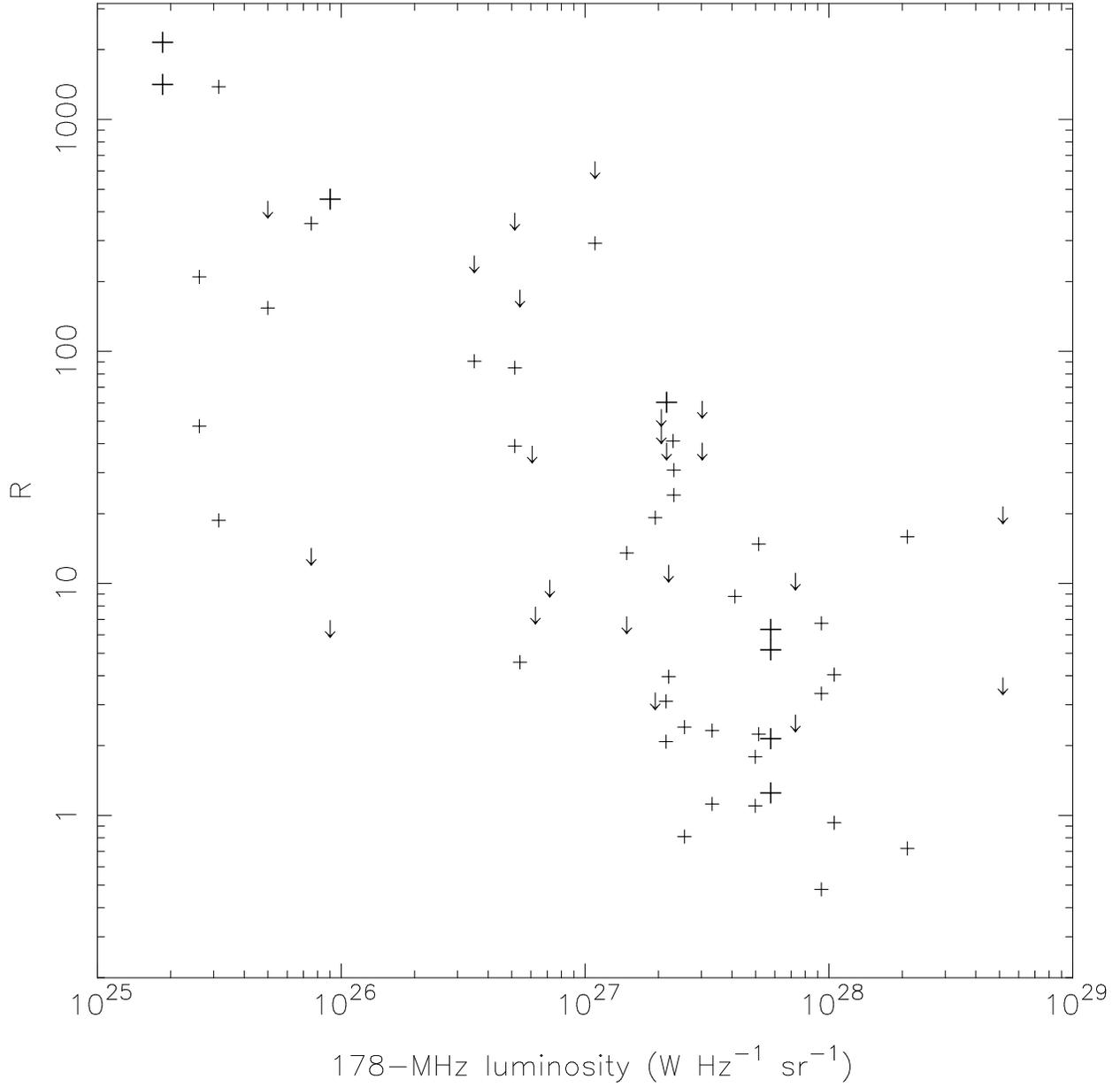}
\end{center}
\caption{$R$ plotted against the total rest-frame 178-MHz luminosity,
  from the 3C/3CRR measurements. Since most sources have more
  than one hotspot, typically two $R$ values are plotted for a given
  source luminosity.}
\label{s178r}
\end{figure*}

\begin{figure*}
\begin{center}
\plotone{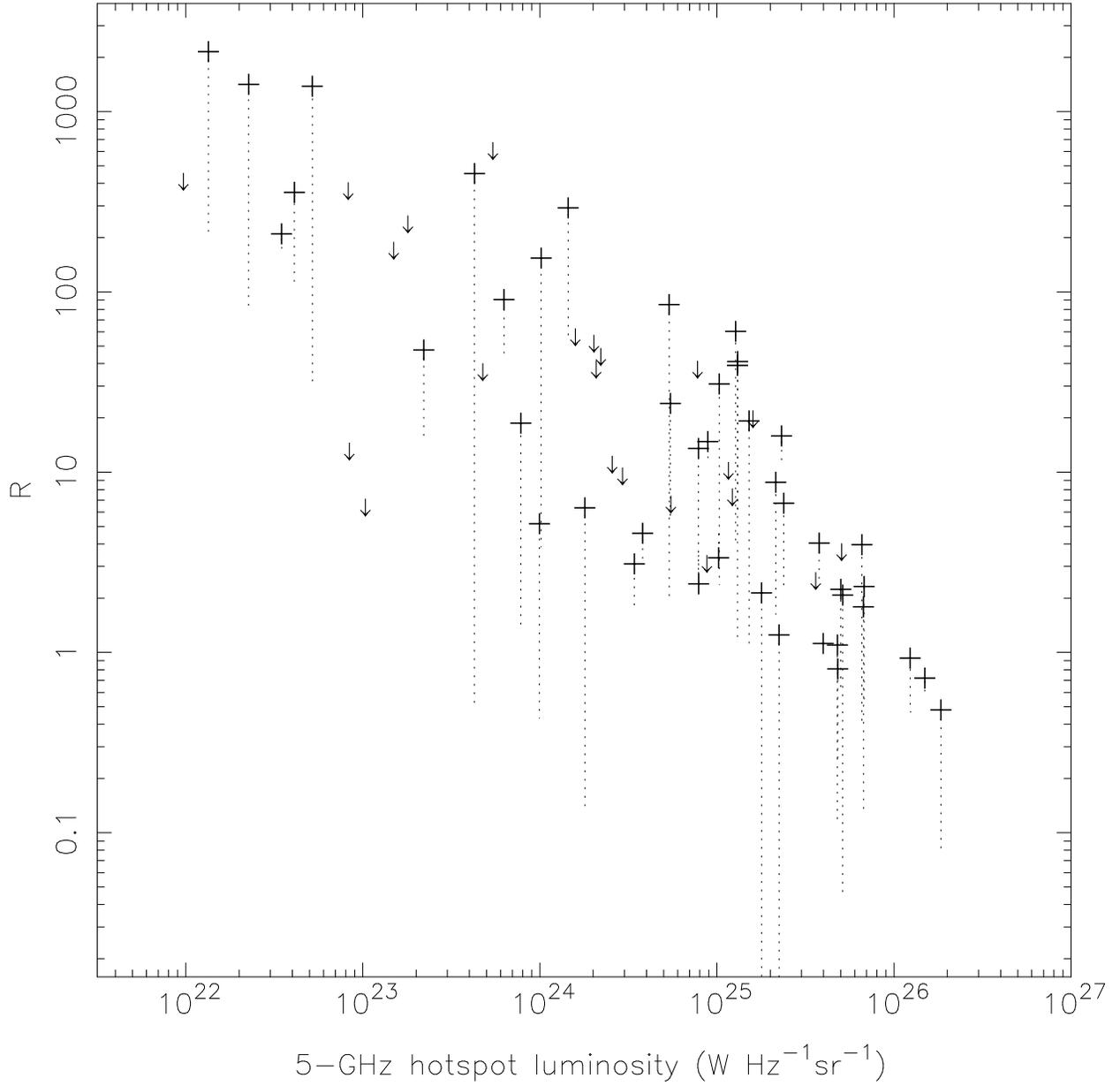}
\end{center}
\caption{$R$ plotted against the rest-frame 5-GHz luminosity of
  the hotspot. The
  dotted lines extending down from the data points show the
  approximate lowest value of $R$ that could have been detected with
  the data, assuming a nominal {\it Chandra} sensitivity of 0.1 nJy at
  1 keV.}
\label{hslr}
\end{figure*}

\begin{figure*}
\plottwo{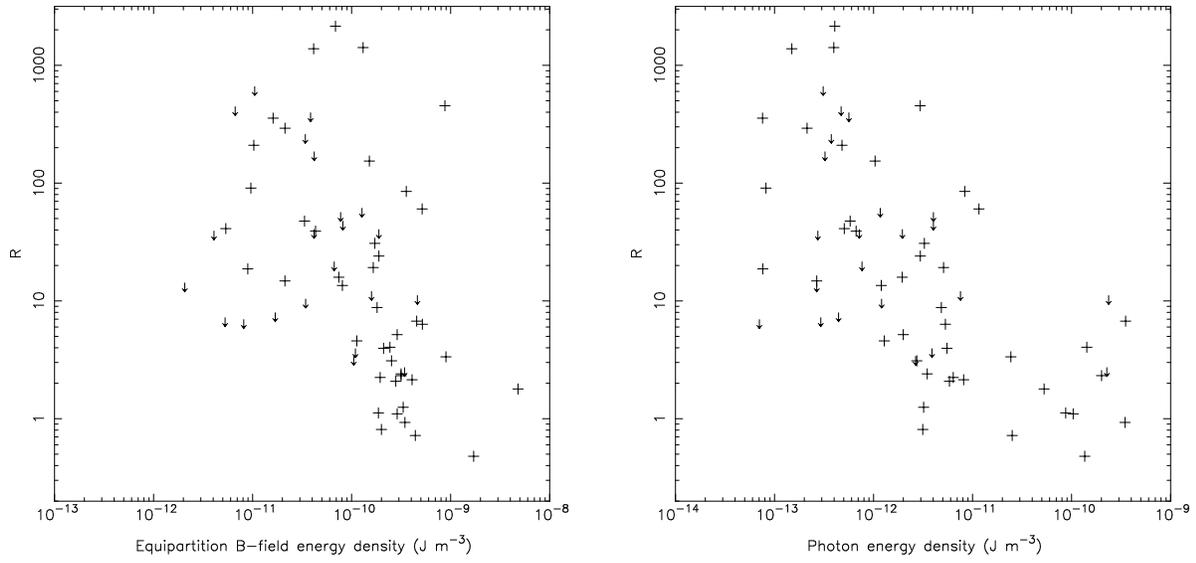}{f6b.eps}
\caption{$R$ plotted against the magnetic field and
  photon energy densities in the hotspots. The plotted photon energy
  density takes into account both synchrotron and CMB photons. Note
  that the two plots have different scales on the $x$-axis.}
\label{energies}
\end{figure*}

What determines the value of $R$ for a particular hotspot? We noted
that the early detections of SSC emission, such as Cygnus A, 3C\,295
and 3C\,123, were all in luminous sources, while well-studied
problematic sources such as 3C\,390.3 and Pictor A are much lower in
overall radio luminosity. Accordingly, we looked for a relationship
between $R$ and total 178-MHz luminosity from the original 3C
measurements or the revised values of \citet{lrl83} (correcting
by a factor 1.09 so as to bring the flux densities on to the scale of
\citet{bgpw77}, and using low-frequency spectral indices to correct
to the rest frame), obtaining the plot shown in Fig.~\ref{s178r}. The
inverse correlation seen here appears to indicate a role of the source
luminosity in determining $R$. The correlation is improved if we plot
the luminosity of only the hotspot against $R$ (Fig.\ \ref{hslr});
here we have used measurements from the radio maps, correcting to a
rest-frame frequency of 5 GHz by assuming a radio spectral index of
0.5. The improvement suggests that the relationship with hotspot
luminosity is primary, and that the correlation with overall source
luminosity arises because of the correlation between hotspot and
source luminosity. It is important to realise that these plots are not
necessarily an indication of a one-to-one correlation between $R$ and
hotspot or total source luminosity. Firstly, {\it Chandra}'s
sensitivity (around 0.1 nJy at 1 keV for the exposure times used in
these observations) means that we would not expect to detect the SSC
emission from the lowest-luminosity hotspots, so that it is
observationally impossible to populate the bottom left-hand corner of
Fig.\ \ref{hslr}, as shown by the dotted lines illustrating the
observational limits. Secondly, there is a positive correlation between
the predicted SSC flux density and the hotspot luminosity, since the
sample is flux-limited and the SSC luminosity is a non-linear function
of the hotspot luminosity (for a given hotspot size) and this
increases the strength of the apparent correlation. However, there
is at least one key result from this analysis; {\it there are no
hotspots with high luminosity and high $R$} --- we would certainly
have been able to detect such hotspots if they existed. By contrast,
low-luminosity hotspots appear to be able to have extremely high $R$
values, though we cannot say definitely that all of them do. We find
no other relationships between $R$ and hotspot or source parameters
such as hotspot angular or linear size, source size, redshift, or
radio spectral index. However, there are relationships between $R$ and
other derived quantities such as equipartition magnetic field energy density
and photon energy density (Fig.\ \ref{energies}); these are not
surprising, since all of the quantities are related to radio
luminosity. We return to the possible physical significance of these
relationships below (\S\ref{synch}).

\begin{figure*}
\begin{center}
\plotone{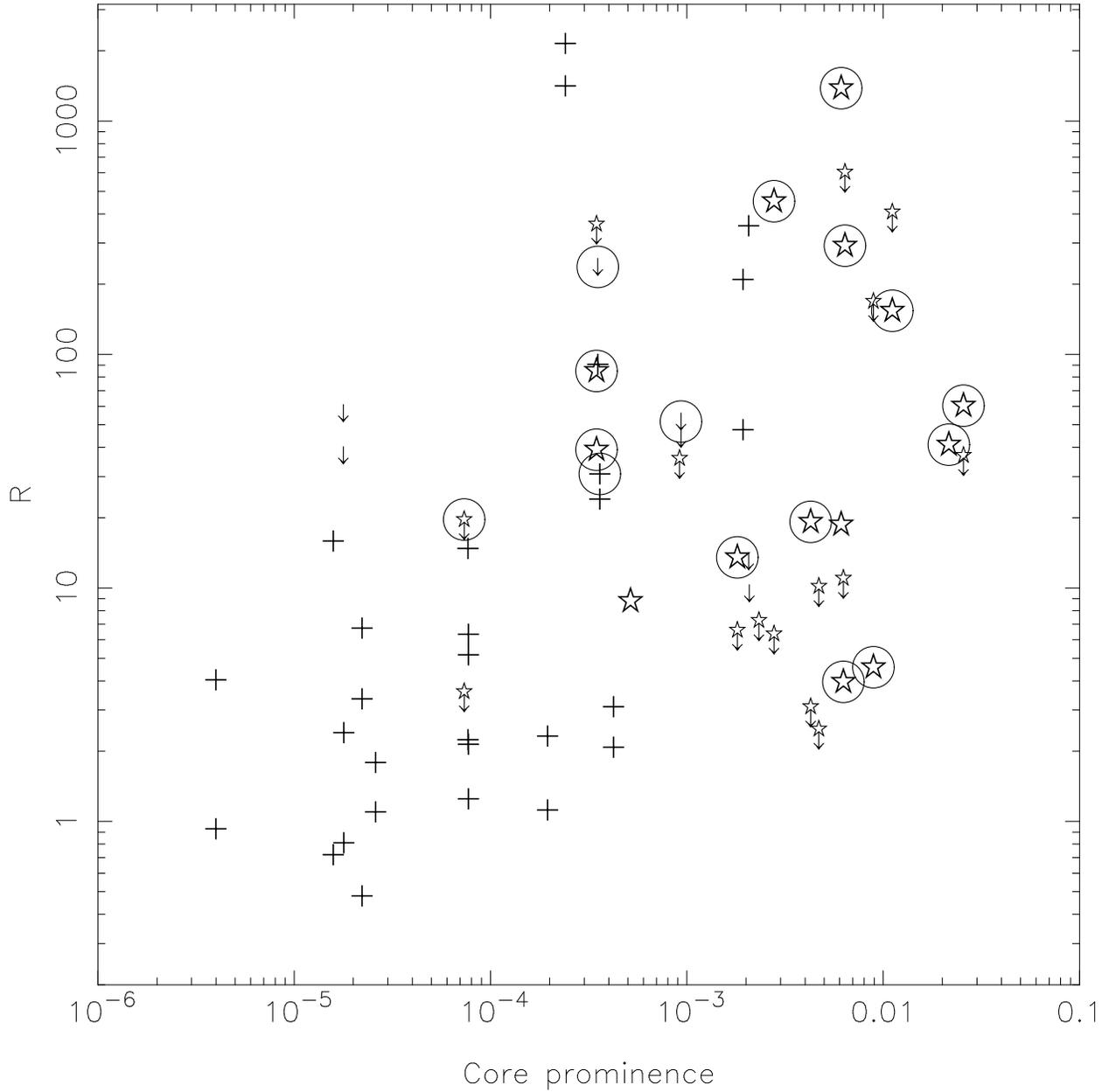}
\end{center}
\caption{$R$ plotted against core prominence. Stars indicate
  broad-line objects (broad-line radio galaxies and quasars) that are
  expected to lie at angles $\la 45^\circ$ to the line of sight in
  unified models.  Circles around
  data points indicate hotspots on the same side of the source as a
  distinct one-sided radio jet.}
\label{hscp}
\end{figure*}

\begin{figure*}
\begin{center}
\plotone{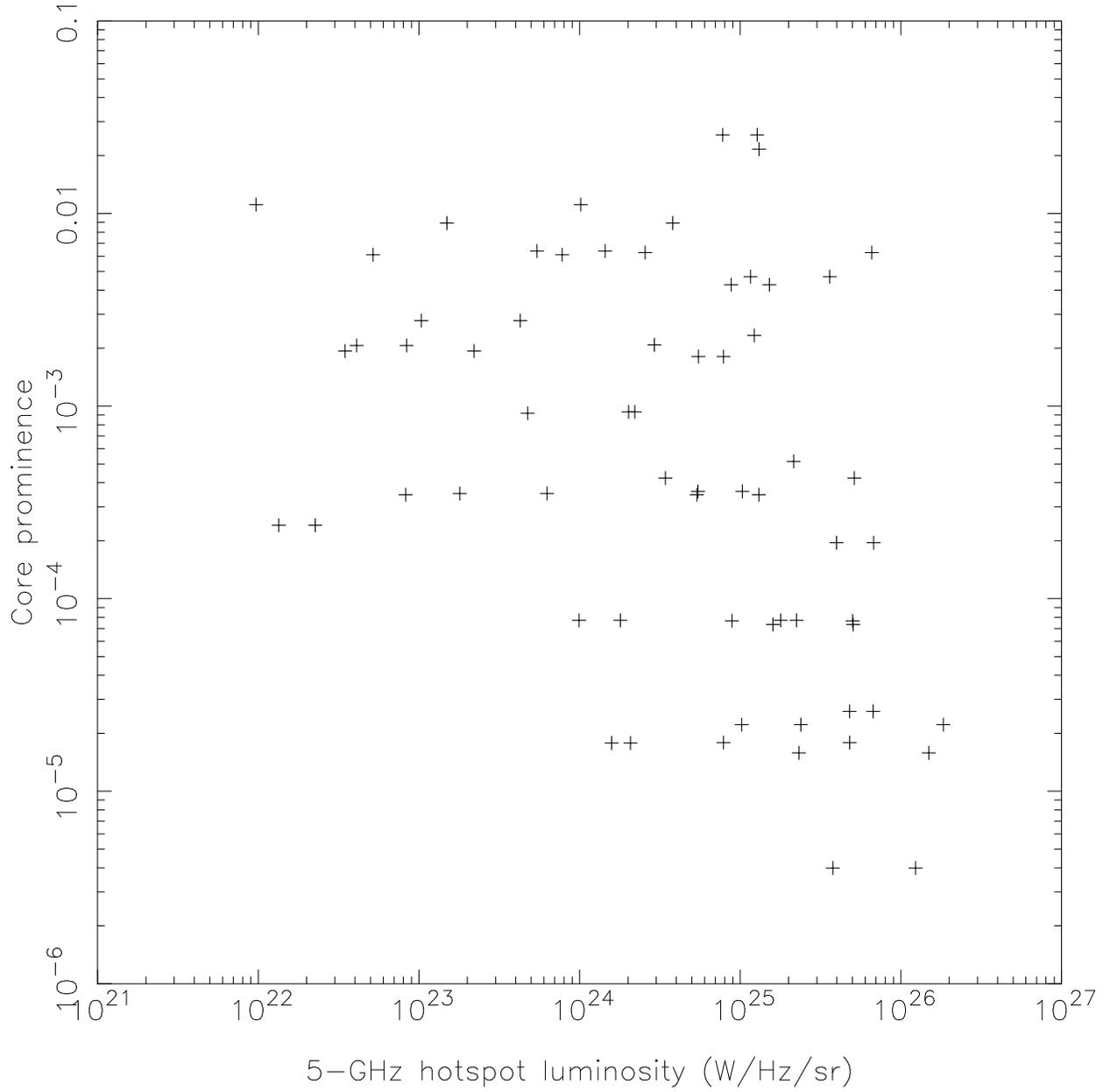}
\end{center}
\caption{Core prominence plotted against hotspot luminosity for the
X-ray hotspot sample.}
\label{cphsl}
\end{figure*}

Finally, we investigated the role of beaming in determining the X-ray
brightness of hotspots by plotting the $R$ parameter against the core
prominence, defined here as the ratio of 5-GHz core flux density to
(rest-frame) 178-MHz total source flux density. Core prominence is
often used as a proxy of beaming
\citep[e.g.][]{ob82,km90,mort97,hapr99}, relying on the assumption
that the {\it intrinsic} fraction of the radio source flux emitted by
the core is similar in all sources and that the observed variation in
core prominence arises from relativistic beaming in the parsec-scale
jet. Using the core flux densities tabulated in Table \ref{obs}, we
produced the plot shown in Fig.~\ref{hscp}. This figure certainly
shows a trend, in the sense that (as was already clear) many of the
X-ray over-bright hotspots are in beamed sources, and often on the
same side as a known one-sided radio jet. At the same time, there is
clearly a good deal of scatter in any correlation --- up to 2.5 orders
of magnitude separate sources with similar core prominences --- and
there are sources that do not fit it at all, such as the low
core-prominence, narrow-line source 3C\,403 ($R>1000$). If we plot
core prominence against hotspot radio luminosity (Fig.\ \ref{cphsl})
we see that there is a tendency for sources with high core prominences
to have low-luminosity hotspots, a trend that can be explained
entirely in terms of a bias towards broad-line radio galaxies and
quasars at low redshifts in the parent sample, so that it is not clear
that the trend seen in Fig.~\ref{hscp} is meaningful. Although the
quasars and broad-line radio galaxies (in unified models, the sources
that should be most strongly affected by beaming) tend to lie at the
upper edge of the envelope of $R$ values for a given luminosity range,
the dominant effect is the luminosity dependence. A partial Kendall's
$\tau$ analysis taking into account the upper limits, performed
according to the prescription of \citet{as96} and using their code,
shows that the correlation between $R$ and core prominence is not
significant at the 95\% confidence level if the luminosity correlation
is taken into account, while the correlation between $R$ and
luminosity is significant even given the core prominence relation. So,
with the current data, we have no significant evidence for a
relationship between hotspot $R$ value and beaming, and the apparent
correlation of Fig.\ \ref{hscp} must be regarded as suggestive at
best.

\section{Discussion}

What models can account for these observations? We examine several in
turn.

\subsection{SSC with luminosity-dependent departures from
  equipartition}

One obvious possibility is that the high $R$ values reflect a
significant departure from equipartition; the belief that
equipartition fields exist in hotspots is, after all, a result of the
study of the most luminous hotspots (chosen because of their high
predicted SSC flux densities). The departure from equipartition in
terms of the ratio of equipartition to true magnetic field strengths,
$B_{\rm eq}/B$, is approximately $R^{0.6}$ , so that the
magnetic field strength would have to be a factor $\sim 100$ lower
than the equipartition value in the hotspots with the highest $R$
values.

We can rule out a simple and attractive model in which such low
magnetic fields account entirely for the luminosity-$R$ correlation.
In this toy model, all hotspots have similar numbers of electrons and
sizes. The radio luminosity $L_R$ goes as $B^{1+\alpha}$, and the IC
luminosity scales in the same way (so long as the synchrotron photon
field remains dominant). But the equipartition prediction for SSC
decreases more rapidly, since this depends on the equipartition
estimate of electron density, which goes approximately as $L_R^{4/7}$
(Appendix B), as well as linearly on the observed photon density. So
we should find that $R \propto L^{-4/7}$, which is not far from the
observed slope. We should also find that the hotspot radio luminosity
scales approximately as $B_{\rm req}^{1.5}$, where $B_{\rm req}$ is
the magnetic field strength required to produce the observed X-ray
emission by inverse-Compton processes, and this is also just about
consistent with the data for our sources. However, we would not expect
to see the observed correlation between 178-MHz {\it total} flux
density and $R$ (Fig.\ \ref{s178r}) in this picture, unless the same
ratio between the true and equipartition magnetic fields persisted
throughout the source. Very low fields in lobes are incompatible with
observations of lobe inverse-Compton emission, among other things; we
see high-$R$ hotspots in sources whose lobes clearly do not show the
same ratio between the observed and predicted inverse-Compton emission
from CMB photon scattering (e.g. 3C\,403, R.\ Kraft \etal\ in preparation).

More generally, there are several arguments that seem to us to
disfavor an SSC model (to be accurate, SSC plus inverse-Compton
scattering of the CMB) with a larger departure from equipartition in
lower-power sources:

\begin{itemize}
\item We know (see \S 1) that some high-$R$ hotspots' X-ray spectra and/or
spatial properties are inconsistent with a pure SSC model (the spectra
of the best-studied high-$R$ sources are all found to be steep,
$\alpha_{\rm X} \approx 1.0$).
\item In a few well-resolved cases (e.g. 3C\,351, H02) the {\it local}
  value of $R$ would be even higher in places than the integrated
  value we quote, representing an even greater challenge for SSC.
\item The fact that a synchrotron model can be fitted through the radio,
optical and X-ray points in some high-$R$ sources would have to be a
coincidence in an SSC model, although this is a weak
constraint; the optical emission might also be SSC, as it is thought
to be in a couple of low-$R$ sources \citep{h01,b02}.
\item An SSC model cannot explain the effects of beaming, if these are
  real; beaming {\it suppresses} SSC emission, so that we should see
  a weak anti-correlation with proxies of beaming like core
  prominence, at least for hotspots on the jet side.
\item The special status of $R=1$ suggests that there are few hotspots
  with $B>B_{\rm eq}$: it is not obvious why the departures from
  equipartition should all be in the sense $B<B_{\rm eq}$.
\item There is no obvious mechanism that fully explains the observed
  luminosity dependence of $R$.
\end{itemize}

\subsection{Luminosity-dependent synchrotron emission}
\label{synch}
A synchrotron model is a good fit to the overall spectrum of some of
the most extreme sources, such as 3C\,390.3 and 3C\,403; if we accept
that X-ray synchrotron emission is possible in some hotspots, as it
certainly is in the jets of FRI sources \citep[e.g.,][]{hbw01b},
then it may contribute to many of them. If the ability of a hotspot to
produce X-ray synchrotron emission depended on its luminosity, then it
might be the case that all hotspots have inverse-Compton emission at a
level consistent with $R=1$ and equipartition magnetic fields, but
that the low-luminosity hotspots have an additional synchrotron
component that may greatly exceed the inverse-Compton emission.

It has already been argued in studies of {\it optical} synchrotron
hotspots \citep[e.g.,][]{myr97,bmpv03} that the high-frequency break
in the synchrotron spectrum is a function of hotspot luminosity, in
the sense that optical emission is much commoner from hotspots of low
radio luminosities. This fact can be explained \citep{bmpv03} in terms
of the lower synchrotron loss rates in the lower magnetic fields
(assuming equipartition) and lower photon densities in hotspots of
lower radio luminosity; we have already seen (Fig.\ \ref{energies})
that there is a correlation between these quantities and $R$ in our
objects. In a standard hotspot spatial/spectral model
\citep[e.g.,][]{hm87} the {\it break} in the synchrotron spectrum
comes about when we average over both the acceleration region itself
and the regions downstream of it, in which synchrotron and
inverse-Compton losses have had time to have an effect; we would not
expect to see a break if we could resolve the acceleration region from
the downstream emission. The high-frequency {\it cutoff} in the
synchrotron spectrum is a direct indicator of physics in the
acceleration region, and results from inefficiency in particle
acceleration at high energies: most importantly, from our point of
view, particle acceleration will become inefficient if the energy loss
timescale in the acceleration region (due to synchrotron and
inverse-Compton losses) becomes shorter than the acceleration
timescale. The fact that essentially all our X-ray data points fall on
or below the line of an extrapolation from the radio with $\alpha_{\rm
RX} = 1.0$ (\S3) shows that any luminosity-dependence of synchrotron
radiation in the X-ray hotspots cannot simply be an effect of a
changing frequency of the spectral break, as \citet{bmpv03} argued for
optical hotspots; $\alpha = 1.0$ is the canonical spectral index above
the break, so that the break alone cannot produce the effect we see.
This interpretation is supported by observations of well-studied
luminous hotspots such as 3C\,405's, where the overall spectrum
requires the {\it cutoff} to be below the X-ray region. We conclude
that the luminosity dependence of $R$ must be an effect of the
synchrotron cutoff.

\begin{figure*}
\begin{center}
\plotone{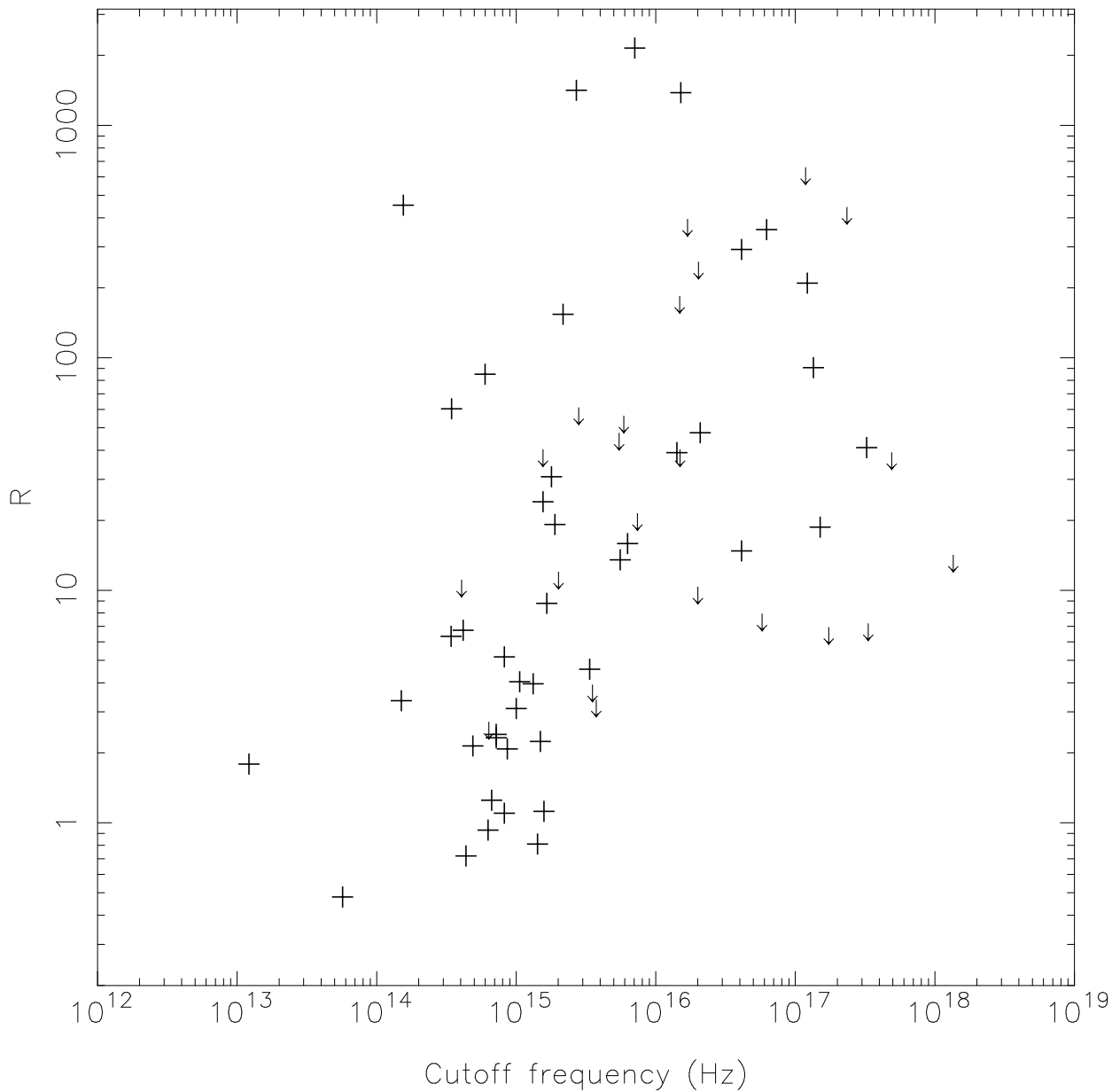}
\end{center}
\caption{$R$ plotted against the synchrotron cutoff frequency for the
  hotspots in the sample, using the proportionality between $\nu_c$
  and magnetic field strength for a constant diffusion coefficient
  quoted in the text. The constant of proportionality (i.e. the
  normalization of the X-axis) is chosen simply to illustrate that the
  magnitude of the effect could be significant in this situation, and
  has no physical basis. The magnetic field strength used is the
  equipartition field derived from our models, and the photon energy
  density is a combination of the microwave background and the
  integrated synchrotron spectrum for each hotspot. A trend is
  apparent, in the sense that low-$R$ hotspots have low cutoffs while
  high-$R$ ones have high cutoffs, but the scatter is large.}
\label{Rnuc}
\end{figure*}

A full calculation of the frequency of the cutoff $\nu_c$ depends on
poorly known quantities such as the magnetic field strength in the
acceleration region and the diffusion coefficient of relativistic
particles. In the simplest case, with uniform magnetic field
throughout the hotspot, a diffusion coefficient independent of both
magnetic field and electron energy, and a non-relativistic shock, it
can be shown that $\nu_c \propto B/({2\over3}B^2 + B_{\rm IC}^2)^2$,
with the constant of proportionality depending on the numerical value
of the diffusion coefficient, where $B$ is the magnetic field strength
and $B_{\rm IC}$ the equivalent inverse-Compton field strength,
defined as $B_{\rm IC} = \sqrt{2\mu_0 U_{\rm IC}}$, with $U_{\rm IC}$
being the energy density in all photon fields. $\nu_c$ in this
calculation does exhibit a change over the parameters of the hotspots
we have studied that would be sufficient in magnitude to explain the
observed effect, although the detailed correlation with $R$ is not
particularly good (Fig.\ \ref{Rnuc}). \citet{bmpv03} consider two
cases with less simplistic forms of the diffusion coefficient
(Kolmogorov and Bohm diffusion coefficients), and are able to
calculate corresponding numerical values for $\gamma_{\rm max}$ (their
eq. 5). The functional form of $\gamma_{\rm max}$ for the Bohm
coefficient means that the cutoff {\it frequency} is essentially
constant, and (for the values quoted by Brunetti \etal) lies well
above the X-ray region for all our hotspots; the Kolmogorov
coefficient gives a $\nu_{\rm c}$ lying well below X-ray frequencies
for equipartition magnetic field strengths in all the hotspots. This
illustrates the strong dependence of the expectation on the unknown
microphysics of the acceleration process. The basic principle of this
model remains plausible: hotspot luminosity (and therefore magnetic
field and photon energy density) are controlling the high-energy
cutoff of the synchrotron spectrum.

Any relationship between $R$ and core prominence is not easy to
explain in a model where much of the X-ray emission is synchrotron.
$R$ does not have a simple dependence on beaming parameters in this
model: the expected synchrotron flux of the hotspot increases with
beaming, but so will the predicted inverse-Compton flux density (from
both SSC and CMB scattering), since the prediction we make is based on
the {\it observed} radio flux density and takes no account of beaming.
We calculated the expected variation of $R$ with angle to the line of
sight $\theta$ for a source whose intrinsic (rest-frame) properties
were held constant. For modest beaming factors, corresponding to $v/c
\sim 0.3$, we find that $R$ does indeed increase as $\theta$ gets
smaller, but only for extremely low-luminosity hotspots, in which
scattering of CMB photons is the dominant IC process; for more
luminous hotspots the trend is reversed, and we would expect $R$ to be
largest for hotspots that are beamed away from us (that is, on the
{\it counterjet} side of beamed sources). In any case, the amount of
variation introduced by this process into the $R$ value distribution
is small, no more than a factor 2 between minimum and maximum values,
for $v/c \sim 0.3$. Much higher speeds ($v/c \ga 0.9$) would be
required, for reasonable hotspot luminosities, to obtain the order of
magnitude scatter in the $R$ parameter (after accounting for the
luminosity dependence) that appears to be present in Fig.\ \ref{hslr}.

\subsection{Deceleration and beaming}
\label{dcb}
We have already shown (\S\ref{model}) that the apparent relationship
between proxies of beaming, such as core prominence, and the X-ray
brightness of the hotspot (H02; \citealt{gk03}) may at
least partly be a selection effect in the available X-ray data. The
current sample of X-ray hotspots, at least at low hotspot
luminosities, is strongly biased towards beamed objects, while beamed
objects are known to have brighter, more compact, flatter-spectrum
hotspots on the jet side. It is now known that there are narrow-line
radio sources that should lie close to the plane of the sky that have
high $R$, and a few examples of sources (for example, 3C\,228 and
3C\,321) where hotspots on both sides of the source have high $R$. It
is not clear, therefore, whether there is any beaming effect that
needs to be explained from an X-ray perspective.

If there {\it is} any beaming effect, then the two models discussed so
far both have difficulty in explaining it, so alternative models must
be considered. The standard way of explaining the (effectively) high
$R$ values in the X-ray jets of core-dominated quasars is to invoke
highly relativistic bulk speeds and the consequent boost of the energy
density of the CMB in the rest frame of the jet. We regard this model
as untenable in the present case, for several reasons. Firstly, large
bulk Lorentz factors are required for even moderate $R$ values (see
the discussion of 3C\,351 in H02) and these in turn constrain the
source to lie at a small angle to the line of sight; this cannot
possibly be the case for all or even most of our high-$R$ objects,
which are drawn from a low-frequency-selected, lobe-dominated sample,
and which include objects that, in unified models, must be close to
the plane of the sky. Secondly, in the standard picture, the radio
emission from the hotspots comes from the post-shock region, and so
high bulk Lorentz factors are hard to achieve; although there are some
effects that are best explained by moderate relativistic beaming in
the post-shock flow, bulk Lorentz factors $\sim 10$ have never been
required by observation, and are in fact inconsistent with the known
properties of hotspots.

A more viable model involving beaming effects is that of \citet{gk03}
(see \S\ref{intro}). The picture they describe almost certainly has to be
true at some level, but a quantitative test is difficult, since it
relies on knowledge of the velocity and electron density structure of
the jet upsteam of the hotspot that is hard to obtain observationally.
In addition, this model cannot account for all the features of our
data, such as the hotspot luminosity dependence of $R$. However, our
data clearly do not rule out a beaming effect at some level and, if it
is present, the standard inverse-Compton or synchrotron models cannot
account for it without involving large speeds. A full test of this
type of beaming model must await an unbiased sample of hotspots in
which orientation and luminosity effects can be clearly separated.

\section{Hotspots and jet knots}
\label{jk}

As discussed in \S1, we have tried to distinguish between hotspots,
defined as structures where the well-collimated flow of the jet
terminates, and jet knots, where the assumption is that the jet
continues more or less unaffected by whatever process produces the
increase in synchrotron emissivity. The key physical differences
between the two systems are (1) that there is probably not a strong
shock in FRII jet knots, since there is little evidence that the jets
decelerate there, and (2) that the particles in jet knots
probably have a shorter dwell time in the region of interest, since
the downstream flow speed is likely to be faster, which could lead to
spectral differences even if the acceleration processes are similar.
In practice the distinction between the two types of feature is
difficult to draw observationally: there are several features that we
have considered to be hotspots in our sample (e.g., 3C\,390.3 N,
3C\,403 F6, 3C\,275.1 N) that might well be jet knots in which the
continuing jet is poorly defined. Equally, it must be the case that
there is continued collimated flow out of primary hotspots in cases
where there is optical or possible X-ray synchrotron emission in the
secondary hotspot, requiring {\it in situ} particle acceleration --
3C\,351's hotspots J and K are a good example. We see no observational
differences between these borderline jet knot/hotspot sources and
clearly defined terminal hotspots.

Should we therefore try to apply the results of the present work to
jet knots as well as to hotspots? As we have argued above
(\S\ref{dcb}) the generally favored jet X-ray emission mechanism for
core-dominated quasars cannot apply to more than a small subset of our
sources, and particularly not the narrow-line objects, some of which
exhibit either possible jet-related X-ray knots (e.g. 3C\,403) or
clear jet-related X-ray emission (e.g. 3C\,321 and 3C\,452 in Appendix
A). The X-ray emission mechanism here seems likely to be synchrotron,
as in the jets of low-luminosity FRI sources. A full analysis of the
known FRII jet-related X-ray emission is beyond the scope of the
present paper, but from our work on hotspots we can make the
`prediction' (borne out by the obserations that we are currently aware
of) that jet-related X-ray synchrotron emission in FRIIs will be seen
mostly in low-luminosity jet knots, and therefore should be
particularly easy to find in low-luminosity FRII sources. The hotspot
behaviour is also qualitatively similar to what is seen in the jets of some
powerful quasars, such as 3C\,273 \citep{sutm01, mhgd01}, in which the
X-ray-to-radio ratio of jet knots decreases as the knot radio flux
density increases.

\section{Conclusions}

We have shown that the properties of the X-ray emission of hotspots
depend strongly on their overall radio luminosity. High-luminosity
hotspots, of the type originally examined in inverse-Compton studies,
consistently show X-ray emission that is close to being consistent with the
predictions of a synchrotron self-Compton model with an equipartition
magnetic field. Low-luminosity hotspots sometimes (and maybe always)
have X-ray emission that is much brighter than would be expected in
this model. We argue that:

\begin{itemize}
\item The good agreement between IC models and data seen for the
  luminous hotspots continues to suggest that the X-ray emission
  mechanism in these systems really is synchrotron self-Compton, that magnetic
  fields really are in equipartition, and that populations of protons
  which dominate energetically by large factors ($\ga 100$) and/or
  very small filling factors are not present.
\item Models in which the unexpectedly strong X-ray emission from some
  low-luminosity hotspots indicate a large departure from
  equipartition in these objects are not plausible for a number of
  reasons: synchrotron emission is more likely.
\item If a synchrotron model is adopted, the high-frequency cutoff of the
  synchrotron spectrum must be dependent on luminosity in order to
  explain the X-ray emission from all hotspots. This is
  physically plausible, but a fully quantitative test depends on the
  microphysics of the acceleration process.
\item There is little significant evidence that relativistic beaming
  is important in the current sample; an unbiased sample of X-ray
  hotspots would be of great importance in testing beaming models.
\item It may be possible to extend our conclusions on hotspots to the
  jet-related X-ray features seen in a number of FRII sources,
  particularly those at relatively large angles to the line of sight;
  if so, we would expect that they would show the same luminosity
  dependence, in the sense that only low-luminosity jets would show
  strong X-ray synchrotron emission. A synchrotron origin for the jets in these
  sources would suggest a continuity between their properties and
  those of the lower-power FRIs, for which a synchrotron
  interpretation is well established \citep[e.g.,][]{hbw01b}.
\end{itemize}

\acknowledgments

MJH thanks the Royal Society for a research fellowship. This work was
partially supported by NASA grant GO3-4132X.

We thank C.\ Cheung and an anonymous referee for helpful comments on
the manuscript. We would like to thank those involved in setting up
full online access to the VLA archive, without which this project
would have been impossible, and in particular John Benson for quick
assistance with problems that arose with the online archive in the
early stages of its operation. We thank Rick Perley for supplying
images of the hotspots of Pictor A.

The National Radio Astronomy Observatory is a facility of the National
Science Foundation operated under cooperative agreement by Associated
Universities, Inc.
\clearpage
\begin{deluxetable}{lrrrrrrrrl}
\small
\tablecaption{3C FRII sources observed with {\it Chandra}}
\tablewidth{18cm}
\tablehead{Source&$z$&$S_{178}$&$\alpha_{\rm R}$&Galactic&$S_{\rm core, 5}$&Type&{\it
      Chandra}&Observing&Date\\
&&(Jy)&&$N_{\rm H}$ ($\times 10^{20}$&(mJy)&&obsid&time (s)&observed\\
&&&&cm$^{-2}$)
}
\startdata
3C\,6.1&0.8404& 14.93  &   0.68 &  17.49  &   4.4&N&3009&36492&2002
Oct 15\\
3C\,9&2.012&19.4&1.12&4.11&4.9&Q&1595&19883&2001 Jun 10\\
3C\,47&0.425& 28.78  &  0.98  &5.34    & 73.6&Q&2129&44527&2001 Jan 16 \\
3C\,109&0.3056 &23.54   & 0.85 & 14.13  &  263&B&4005&45713&2003 Mar
23\\
3C\,123&0.2177 &206.01  & 0.70 & 43     &  100&E&829&38465&2000 Mar 21\\
3C\,173.1&0.292&  16.79 &   0.88 & 5.25   &  7.4&E&3053&23999&2002
Nov 06\\
3C\,179&0.846&9.27&0.73&4.32&371&Q&2133&9334&2001 Jan 15\\
3C\,184&0.994&14.39&0.86&3.46&$<0.2$&N&3226&18886&2002 Sep 22\\
3C\,207&0.684 & 14.82   & 0.90 & 5.40  &   510&Q&2130&37544&2000 Nov
04\\
3C\,200&0.458& 12.32  &   0.84 & 3.69 &35.1&N&838&14660&2000 Oct 06\\
3C\,212&1.049 &  16.46  &   0.92 &  4.09&150&Q&434&18054&2000 Oct 26\\
3C\,215&0.411 &  12.43  &   1.06 &  3.75&16.4&Q&3054&33803&2003 Jan 02\\
3C\,219&0.1744 & 44.91  &   0.81 &  1.48&51&B&827&17586&2000 Oct 11\\
3C\,220.1& 0.61 &   17.22  &   0.93 &  1.93  &    25
&N&839&18922&1999 Dec 29\\
3C\,228&0.5524&  23.76 &   1.0 &  3.28  &   13.3&N&2453&13785&2001
Apr 23\\
3C\,254&0.734& 21.69   & 0.96  &1.75    & 19&Q&2209&29668&2001 Mar 26\\
3C\,263&0.652& 16.57  &  0.82 & 0.91   &  157&Q&2126&44148&2000 Oct 28\\
3C\,265&0.8108&  21.26 &   0.96&  2.05  &   2.89&N&2984&58921&2002
Apr 25\\
3C\,275.1&  0.557&  19.95  &  0.96&  1.89 &    130&Q&2096&24757&2001
Jun 02\\
3C\,280&    0.996&  25.83  &  0.81&  1.25 &    1.0&N&2210&63528&2001
Aug 27\\
3C\,281&0.602&6.00&0.71&2.2&19.5&Q&1593&15851&2001 May 30\\
3C\,294&   1.78  & 11.23   & 1.07 & 1.20  &   0.53&N&3207&122020&2002
Feb 27\\
3C\,295&0.4614& 91.02  &  0.63 & 1.38   &  3&N&2254&90936&2001 May 18\\
3C\,303&0.141&   12.21 &   0.76&  1.60  &   150&B&1623&14951&2001
Mar 23\\
3C\,321&0.096&  14.72  &  0.60 & 4.10   &  30&N&3138&47130&2002 Apr 30 \\
3C\,324&1.2063& 17.22  &  0.90 & 4.47   &  $<0.14$&N&326&42147&2000
Jun 25\\
3C\,330&0.5490& 30.30  &  0.71 & 2.94   &  0.74&N &2127&44083&2001
Oct 16 \\
3C\,334&0.555&  11.88  &  0.86 & 4.14   &  111&Q&2097&32468&2001 Aug
22\\
3C\,351&0.371&  14.93  &  0.73 & 2.03   &  6.5 &Q &2128&45701&2001
Aug 24\\
3C\,390.3&0.0569& 51.78  &  0.75 & 3.74   &  330 &B&830&33974&2000
Apr 17\\
3C\,401& 0.201 &  22.78  &   0.71 &  7.42  &    32&E&3083&22666&2002 Sep 20 \\
3C\,403&0.0590&28.3&0.74&13.56&7.1&N&2968&49472&2002 Dec 07\\
3C\,405&0.0565&9660&0.74&33.0&776&N&360&34720&2000 May 21\\
3C\,427.1& 0.572 &  28.99   &  0.97 &  11.60  &
0.8&E&2194&39456&2002 Jan 27 \\
3C\,438&  0.290 &  48.72  &   0.88 &  17.22  &   16.2&E&3967&47272&2002 Dec 27 \\
3C\,452&0.0811& 59.30  &   0.78 &  11.30  &   130&N &2195&79922&2001
Aug 21\\
Pictor A&0.03498&400&1.0&4.2&1150&B&346&25734&2000 Jan 18\\
\enddata
\label{obs}
\tablecomments{$S_{178}$ is the 178-MHz flux density, on the scale of
  \cite{bgpw77}, mostly taken from \citet{lrl83} or
  \cite{sdma85}. $\alpha_{\rm R}$ is the low-frequency
  spectral index, typically between 178 and 750 MHz. Types are based
  on optical and emission-line characteristics, and are as follows: E,
  low-excitation radio galaxy; N, narrow-line radio galaxy; B,
  broad-line radio galaxy; Q, quasar. Values of $S_{\rm core, 5}$, the
  5-GHz core flux density, are mostly taken from the compilation on
  the 3CRR web pages (http://www.3crr.dyndns.org/) or from the radio maps
  referred to in this paper. All {\it Chandra} datasets are from the
  ACIS-S except 3C\,295, where the ACIS-I was used. No grating data
  met our selection criteria. Livetimes are the filtered times if
  filtering was carried out, and the uncorrected livetime otherwise.}
\end{deluxetable}

\begin{deluxetable}{llrrlr}
\tablecaption{VLA radio observations used in this paper}
\tablewidth{15cm}
\tablehead{Source&Proposal ID&Frequency&Time on&Date&Reference\\
&&(GHz)&source (h)&&(if published)}
\startdata
3C\,6.1&AP380&8.5&1&1999 Aug 02\\
&&8.5&1\tablenotemark{b}&2000 Jan 18\\
3C\,9&AB369&4.9&4.5&1986 May 04&5\\
3C\,47&AB796&4.8&7&1996 Nov 07\\
3C\,109&&8.4&&&1\\
3C\,123&&8.4&&&2\\
3C\,173.1&&8.4&&&2\\
3C\,179&AC150&4.9&0.5&1986 Mar 21\\
3C\,184&&4.9&&&3\\
3C\,200&&8.4&&&1\\
3C\,207&AB796&8.5&3&1996 Nov 08\\
3C\,212&AB796&8.5&4&1996 Nov 07\\
3C\,215&&4.8&&&4,5\\
3C\,219&&1.5&&&4\\
3C\,220.1&&8.4&&&6\\
3C\,228&&8.4&&&1\\
3C\,254&AB522&4.9&0.5&1989 Feb 01\\
3C\,263&&4.9&&&5\\
3C\,265&AF186&4.8&3&1990 Apr 22&11\\
3C\,275.1&&8.4&&&1\\
3C\,280&AV157&8.4&0.7&1988 Dec 22\\
3C\,281&AB631&1.4&0.3&1992 Nov 18\\
3C\,294&AM224&4.7&3.5&1987 Oct 11\\
3C\,295&&8.4&&&1\\
3C\,303&KRON&4.9&0.3\tablenotemark{a}&1981 Apr 20&10\\
3C\,321&AV127&4.8&3.7&1986 Apr 10\\
&&1.5, 4.8&2.9, 3.7\tablenotemark{b}&1986 Aug 29\\
3C\,324&AF186&4.9&3&1990 Apr 22&11\\
3C\,330&&8.4&&&1\\
3C\,334&&4.9&&&5\\
3C\,351&&8.4&&&1\\
3C\,390.3&&1.5&&&4\\
3C\,401&&8.4&&&2\\
3C\,403&&8.4&&&7\\
3C\,405&&4.5&&&8\\
3C\,427.1&&8.4&&&1\\
3C\,438&&8.4&&&2\\
3C\,452&&8.4&&&7\\
Pictor A&&4.9&&&9\\
\enddata
\label{radioobs}
\tablecomments{We list VLA observational details only for observations
  that we have retrieved from the archive and reduced ourselves in the
  course of this project; for other observations we were able to
  obtain electronic maps from others (or already had them ourselves)
  and the reader is referred to the references given below for the
  observational information. All data retrieved from the archive were
  taken with the VLA in its A configuration, except where otherwise
  noted.} \tablerefs{(1) \citealt{grhc04}; (2) \citealt{hapr97}; (3)
  \citealt{bwhb04}; (4) \citealt{lbs98} (the 3CRR Atlas); (5)
  \citealt{bhlb94}; (6) \citealt{wbhl01}; (7) \citealt{bblp92}; (8)
  \citealt*{pdc84}; (9) \citealt*{prm97}; (10) \citealt{k86}; (11)
  \citealt{fbbp93}.} \tablenotetext{a}{Only one observing frequency of
  12.5 MHz bandwidth was used.} \tablenotetext{b}{B-configuration
  data.}
\end{deluxetable}

\begin{deluxetable}{llrrrrr}
\tablecaption{Radio and X-ray flux densities and predicted inverse-Compton flux
  densities}
\tablehead{Source&Hotspot&Ang. size&5-GHz radio&1-keV flux&Predicted flux&Ratio $R$\\
&&(arcsec)&flux (Jy)&(nJy)&(nJy)&(observed/\\
&&&&&&predicted)}
\startdata
3C\,6.1&N&0.36&0.340&0.45&0.19&2.3\\
&S&0.41&0.200&0.09&0.081&1.1\\
3C\,9&N&0.38&0.038&$<0.09$&0.024&$<3.6$\\
&S&0.39&0.012&$<0.09$&0.0046&$<20$\\
3C\,47&S&0.434&0.181&0.54&0.040&14\\
&N&1.89&0.127&$<0.1$&0.015&$<6.6$\\
3C\,109&S&0.377&0.181&0.15&0.033&4.6\\
&N&0.274&0.007&$<0.09$&0.00053&$<169$\\
3C\,123&E&$1.1\times0.54$&5.12&4.6&2.2&2.1\\
&W&$1.0\times 0.13$&0.341&0.18&0.059&3.1\\
3C\,173.1&S&0.83&0.033&0.2&0.0022&91\\
&N&0.26&0.009&$<0.12$&0.00051&$<237$\\
3C\,179&W&0.145&0.063&1.54&0.026&63\\
&E&0.45&0.038&$<0.26$&0.0070&$<37$\\
3C\,200&N&0.6&0.057&$<0.1$&0.011&$<9.5$\\
3C\,207&E&0.27&0.044&0.69&0.0095&73\\
3C\,212&N&0.144&0.035&$<0.14$&0.014&$<10$\\
&S&0.25&0.110&$<0.14$&0.056&$<2.5$\\
3C\,215&E&1.0&0.012&$<0.04$&0.0011&$<36$\\
3C\,220.1&E&0.27&0.021&$<0.13$&0.0025&$<52$\\
&W&0.27&0.023&$<0.13$&0.0030&$<44$\\
3C\,228&N&0.203&0.070&0.45&0.019&24\\
&S&0.265&0.132&1.3&0.042&31\\
3C\,254&W&0.29&0.146&0.54&0.061&8.8\\
3C\,263&E&0.39&0.582&1.0&0.25&4.0\\
&W&0.18&0.023&$<0.06$&0.0054&$<11$\\
3C\,265&E&0.356&0.272&0.35&0.16&2.2\\
&W&0.73&0.048&0.13&0.088&15\\
3C\,275.1&N&$1.4\times0.2$&0.191&1.78&0.093&19\\
&S&0.378&0.111&$<0.12$&0.038&$<3.1$\\
3C\,280&E&0.186&0.082&0.31&0.046&6.7\\
&W&0.146&0.631&0.6&1.2&0.48\\
&Wc&0.081&0.035&0.07&0.021&3.35\\
3C\,281&N&1.04&0.129&$<0.16$&0.022&$<7.3$\\
3C\,294&N&0.283&0.143&0.12&0.17&0.72\\
&S&0.43&0.022&0.14&0.0088&16\\
3C\,295&N&0.1&1.29&1.4&0.78&1.8\\
&S&0.1&0.92&0.94&0.85&1.1\\
3C\,303&W&$1.1\times0.28$&0.257&4.0&0.026&154\\
&E&0.57&0.0025&$<0.16$&0.00039&$<408$\\
3C\,321&E&0.69&0.125&0.3&0.006&48\\
&W&$2.7\times 0.45$&0.020&0.12&0.00057&210\\
3C\,324&E&0.365&0.277&0.20&0.21&0.93\\
&W&0.301&0.085&0.16&0.040&4.04\\
3C\,330&N&0.45&0.625&0.35&0.42&0.81\\
&S&0.20&0.102&0.068&0.028&2.4\\
3C\,334&S&$1.34\times0.3$&0.018&0.54&0.0018&292\\
&N&0.5&0.007&$<0.4$&0.00066&$<604$\\
3C\,351&J&0.16&0.167&4.3&0.051&85\\
&K&0.8&0.406&3.4&0.087&39\\
&S&0.16&0.0025&$<0.05$&0.00014&$<362$\\
3C\,390.3&N&$1.3\times 0.5$&0.087&4.5&0.003&1380\\
&S&3.7&1.30&1.3&0.07&19\\
3C\,403&F1&0.275&0.021&1.0&0.00047&2149\\
&F6&0.256&0.035&1.8&0.0013&1414\\
3C\,405&A&$2.5\times1.25$&38.0&19.4&15.5&1.3\\
&B&0.44&3.04&4.5&0.71&6.3\\
&D&1.09&30.3&29.2&13.6&2.1\\
&E&$0.45\times0.63$&1.68&1.2&0.23&5.2\\
3C\,427.1&N&0.165&0.019&$<0.17$&0.0030&$<56$\\
&S&0.14&0.025&$<0.17$&0.0046&$<37$\\
3C\,452&W&0.705&0.033&0.34&0.00095&356\\
&E&3.0&0.067&$<0.05$&0.0039&$<13$\\
Pic A&W&0.75&1.93&89&0.20&454\\
&E&0.75&0.467&$<0.16$&0.044&$<6.4$\\
\enddata
\label{fluxes}
\tablecomments{The hotspot identifier is usually N, S, E or W,
  referring to the obvious or brightest hotspot in the north, south,
  east or west lobes. Exceptions are made where a multiple-hotspot
  source has names for the individual components that are used
  relatively widely in the literature; this is true of 3C\,405
  (notation of \citealt{hr74}), 3C\,351 (notation of \citealt{bhlb94})
  and 3C\,403 (notation of \citealt{bblp92}). The angular sizes quoted
  are the radii of homogeneous sphere models fitted to the radio data,
  as described in the text, except where two numbers are quoted, in
  which case they are the length and radius of a cylinder and are
  generally directly measured from high-resolution maps. The measured hotspot
  radio flux densities have been scaled to a lab-frame radio frequency
  of 5 GHz using a spectral index $\alpha = 0.5$ for ease of
  comparison. The 1-keV flux densities are the values inferred from
  spectral fitting or the observed count rate, as described in the
  text, and are the unabsorbed fluxes (assuming Galactic absorption).}
\end{deluxetable}

\begin{deluxetable}{llrrlrrlrl}
\tablecaption{Optical flux densities used in this paper and associated
{\it HST} observational details}
\tablewidth{18cm}
\tablehead{Source&HS&Freq.&Flux&Origin&Ref.&Obsid&Filter&Time on&Date\\
&&($\times 10^{14}$&density&&&&&source&observed\\
&&Hz)&($\mu$Jy)&&&&&(s)}
\startdata
3C\,47&S&5.5&$<0.8$ ($<1.0$)&{\it HST}&7&U4492101&F555W&600&1999 Jan 30\\
3C\,109&S&4.5&$<0.8$$ (<1.4$)&{\it HST}&7&U27L1S01&F702W&560&1995 Aug 24\\
3C\,123&E&5.5&$<2.3$&{\it HST}&1&U4494801&F555W&600&1999 Apr 05\\
&W&&$<0.45$ ($<1.5$)\\
3C\,173.1&S&4.3&$<1.2$ ($<1.4$)&{\it HST}&7&U27L2O01&F702W&300&1994 Jul 27\\
3C\,179&W&5.5&$<0.46$ ($<0.57$)&{\it HST}&7&U4495C01&F555W&600&1999 Mar 06\\
3C\,207&E&5.5&$<0.35$ ($<0.46$)&{\it HST}&7&U4498701&F555W&600&1999 Jan 18\\
3C\,228&N&3.5&$<0.85$ ($<0.94$)&{\it HST}&7&U6FA3701&F785LP&2000&2001 May 30\\
&S&&1.04 (1.14)\\
3C\,254&W&5.5&$<0.38$ ($<0.41$)&{\it HST}&7&U4490O01&F555W&600&1999 Mar 18\\
3C\,263&E&4.5&0.8 (0.8)&{\it HST}&5&U2SE0201&F675W&1000&1996 Feb 18\\
3C\,265&E&5.5&$<0.4$ ($<0.4$)&{\it HST}&7&U2CT0J02&F555W&1700&1995 Apr 01\\
&W&4.3&$<1.0$ ($<1.1$)&{\it HST}&7&U27L4F01&F702W&300&1995 May 12\\
3C\,275.1&N&4.7&0.44 (0.48)&{\it HST}&7&U2SE0301&F675W&1800&1995 Jul 25\\
3C\,280&E&4.9&0.32 (0.34)&{\it HST}&7&U2GX0801&F622W&8800&1994 Aug 22\\
&W&&$<0.4$ ($<0.4$)\\
&Wc&&$<0.4$ ($<0.4$)\\
3C\,295&N&4.3&0.078 (0.082)&{\it HST}&4&U2C40A01&F702W&12600&1996 Jan 14\\
&S&&0.02 (0.02)\\
3C\,303&W&5.5&7.5&OHP&2\\
3C\,330&N&5.5&$<0.5$ ($<0.6$)&{\it HST}&5&U3A14X01&F555W&600&1996 Jun 03\\
&S&&$<0.5$ ($<0.6$)\\
3C\,334&S&5.5&$<0.7$ ($<0.86$)&{\it HST}&7&U4492V01&F555W&600&1998 Dec 27\\
3C\,351&J&4.3&2.4 (2.6)&{\it HST}&5&U2X30601&F702W&2400&1995 Nov 30\\
&K&&1.9 (2.1)\\
3C\,390.3&N&4.5&2.2 (2.6)&NOT&6&&&&1996 Jul 18\\
3C\,403&F1&4.3&0.66 (1.08)&{\it HST}&8&U27L7601&F702W&280&1994 Jun 26\\
&F6&&1.32 (2.16)\\
3C\,405&A&4.6&$<80$&Calar&2&&&&1985 Oct\\
&B&&$<46$&Alto&2\\
&D&&$<5$&&2\\
Pictor A&W&4.5&130&ESO 3.6m&3&&&&1985 Nov 08\\
\enddata
\label{hst}
\tablecomments{Hotspots are identified as in Table \ref{fluxes}.
    Optical flux densities are corrected for Galactic extinction if
    only one value is given; where two are given the second (in
    parentheses) is the corrected value. {\it
    HST} observational details are given where {\it HST} data were
    used; other data points are taken from ground-based observations
    described in the literature. Data points for the same object have
    the same origin unless different origins are explicitly listed in
    the Table. NOT indicates the Nordic Optical Telescope and OHP the
    Observatoire de Haute Provence.}
\tablerefs{(1) \citealt{hbw01a}; (2) \citealt{myr97};
  (3) \citealt{rm87}; (4) \citealt{hnpb00}; (5)
  H02 (6) \citealt{hll98}; (7) This
  paper; (8) R.\ Kraft \etal\ , in prep.}
\end{deluxetable}
\clearpage
\appendix
\section{Newly detected hotspots}
Below we present images of the hotspots newly detected in the course
of this work that are not expected to be discussed in more detail in
other papers.
\subsection{3C\,6.1}
\begin{figure}
\begin{center}
\plotone{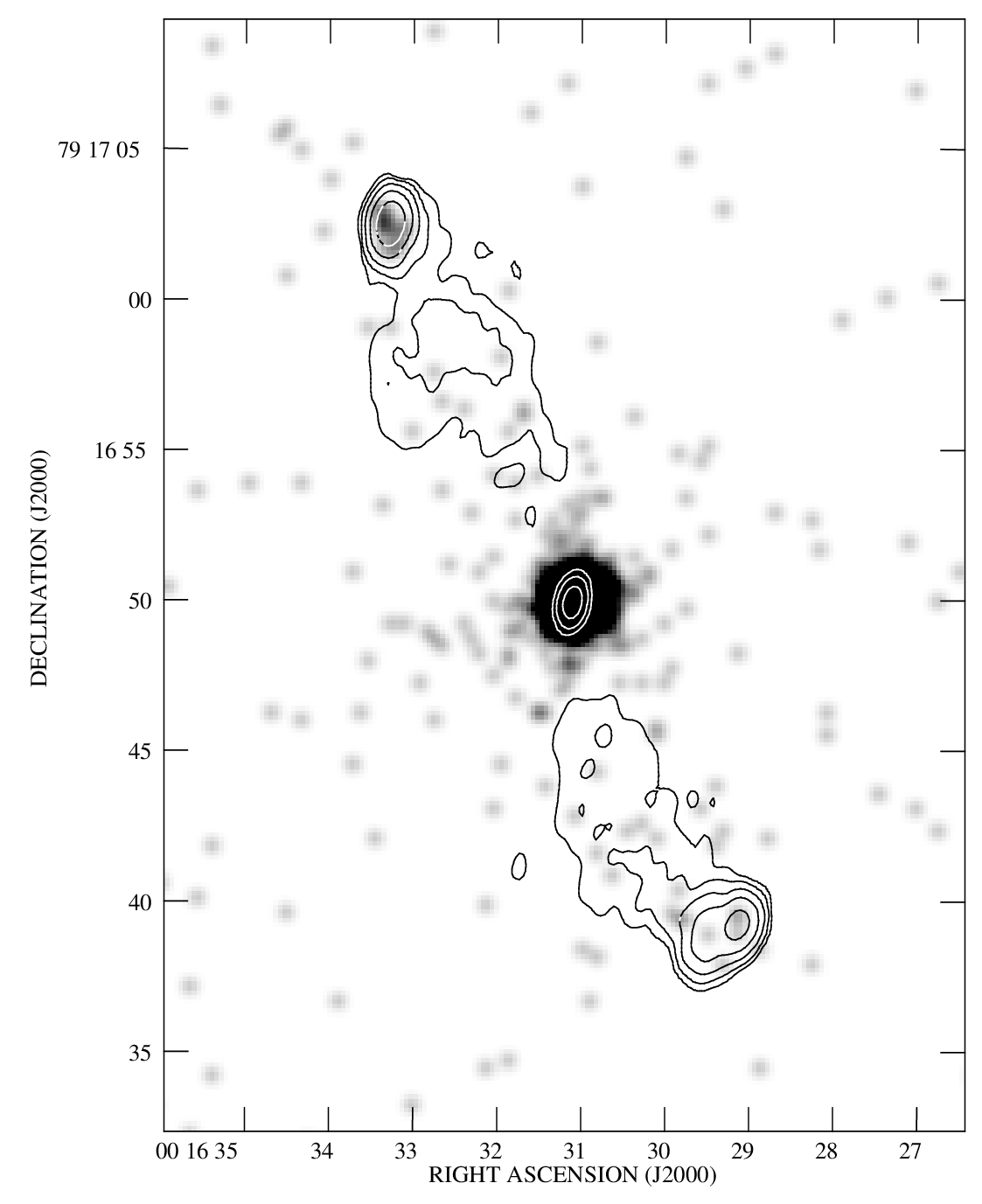}
\caption{The X-ray hotspots of 3C\,6.1. The greyscale shows the 0.5-5
  keV {\it Chandra} counts smoothed with an $0\farcs5$ FWHM Gaussian;
  black corresponds to 1 count per $0\farcs246$ pixel. The contours are of the 8.4-GHz
  B-configuration VLA map at $0\farcs94 \times 0\farcs53$ resolution, and are at $0.2
  \times (1,4,16\dots)$ mJy beam$^{-1}$.}
\label{3c6.1}
\end{center}
\end{figure}

The N hotspot of this narrow-line source is the clearest detection
(Fig.\ \ref{3c6.1}), but there is a weak detection of the S hotspot
too, at well over $3\sigma$ significance. Some X-ray emission is
associated with the lobes.
\subsection{3C\,47}
\begin{figure}
\begin{center}
\plotone{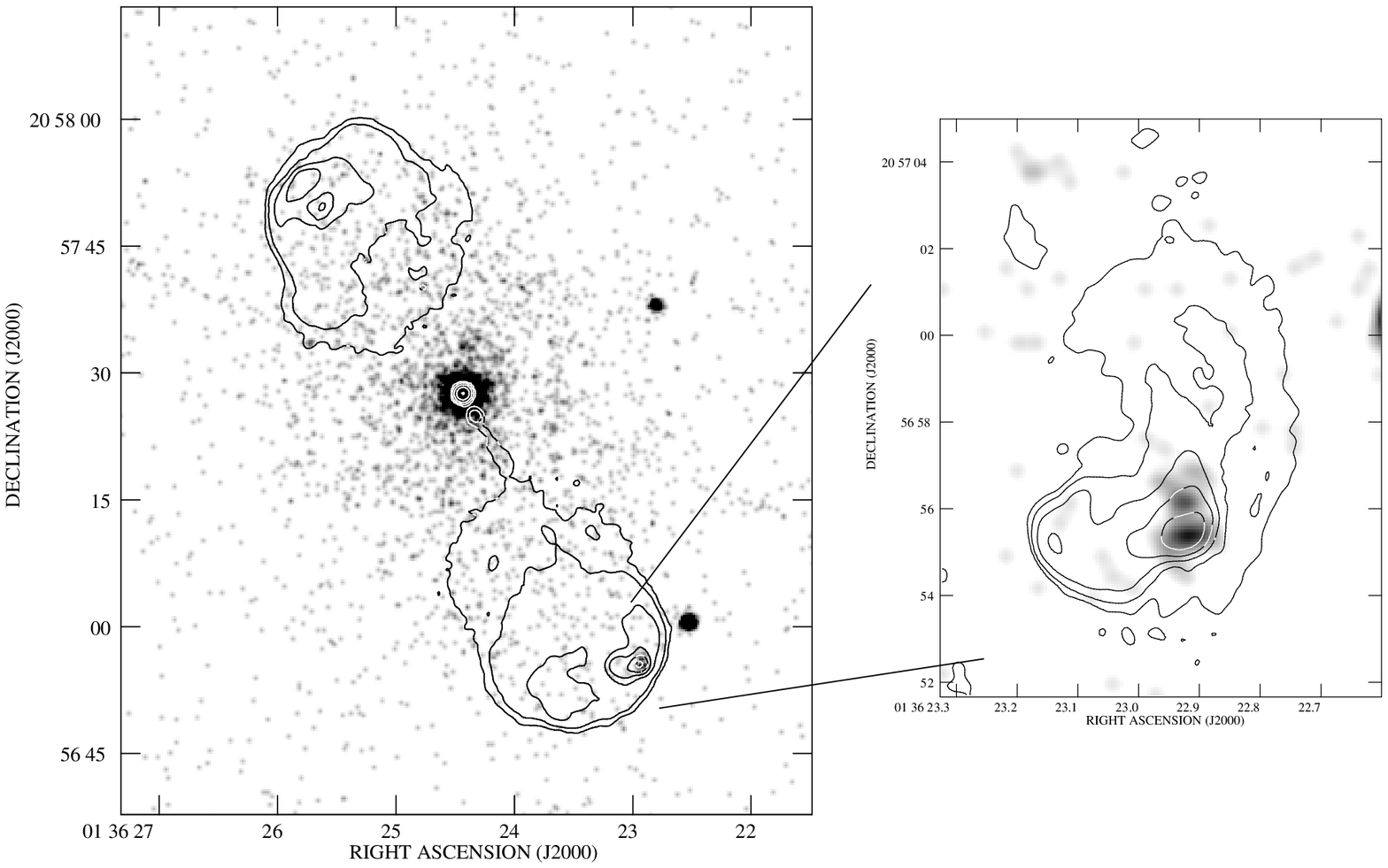}
\caption{The southern X-ray hotspot of 3C\,47. The main greyscale
  (left) shows the 0.5-5 keV {\it Chandra} counts smoothed with an
  $0\farcs5$ FWHM Gaussian; black corresponds to 5 counts. The
  contours are of a $1\farcs0$ resolution 1.6-GHz VLA map taken from
  \citet{lbs98}, and are at $0.3 \times (1,4,16\dots)$ mJy
  beam$^{-1}$. The inset (right) is the same X-ray image with the same
  greyscale level, but with contours from the $0\farcs39 \times
  0\farcs36$ resolution 4.8-GHz VLA map at $80 \times (1,4,16\dots)$
  $\mu$Jy beam$^{-1}$. }
\label{3c47}
\end{center}
\end{figure}

This quasar's bright S hotspot (on the jet side) is detected
(Fig.\ \ref{3c47}), but there is no obvious detection of the fainter N
hotspot. Extended emission is clearly visible in the X-ray image:
since it is extended in the direction of the lobes some of it may well
be inverse-Compton emission, but probably a large fraction of it comes
from a cluster environment, particularly as 3C\,47 exhibits a strong
Laing-Garrington effect \citep[e.g.,][]{l96}.

\subsection{3C\,109}

The southern hotspot of this broad-line radio galaxy is detected; the
southern side is the jet side and a weak jet can be traced into the
hotspot \citep{grhc04}. Some excess extended emission from the lobes
can be seen in these images, and is consistent with inverse-Compton
emission at approximately the level expected from equipartition in the
lobes.

\begin{figure}
\begin{center}
\plotone{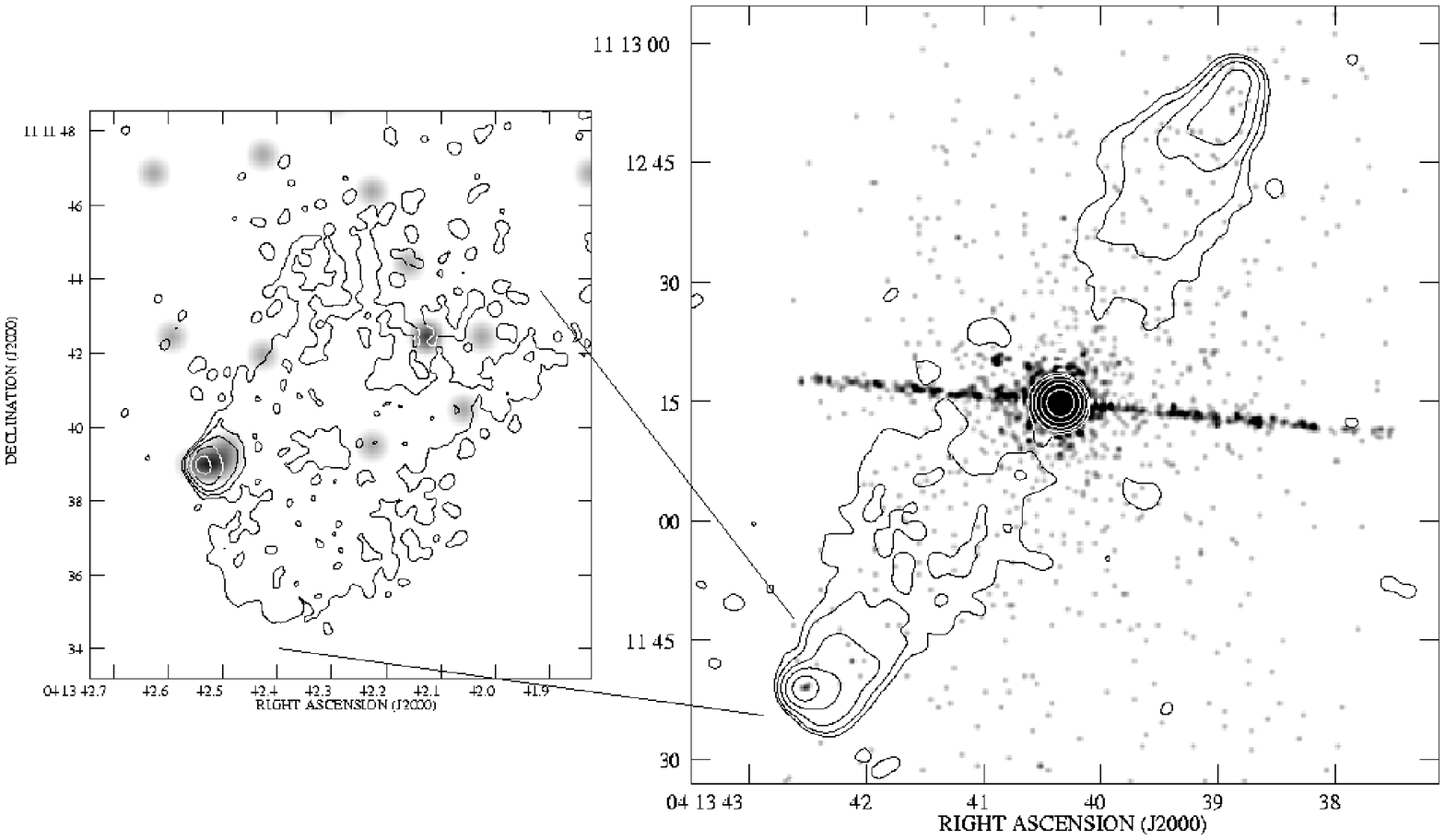}
\caption{The southern X-ray hotspot of 3C\,109. The main greyscale (right)
  shows the 0.5-5 keV {\it Chandra} counts smoothed
  with an $0\farcs5$ FWHM Gaussian; black corresponds to 1.5 counts
  per $0\farcs492$ pixel. The line across the image is the {\it
  Chandra} readout streak. The contours are of a $2\farcs5$ resolution
  8.4-GHz VLA map taken from \citet{grhc04}, and are at $0.2 \times
  (1,4,16\dots)$ mJy beam$^{-1}$. The inset on the left shows the
  hotspot; the X-ray map is the same but contours are from a
  $0\farcs25$ resolution 8.4-GHz map also from \citet{grhc04}, at $0.1
  \times (1,4,16\dots)$ mJy beam$^{-1}$.}
\label{3c109}
\end{center}
\end{figure}

\subsection{3C\,173.1}
\begin{figure}
\begin{center}
\plotone{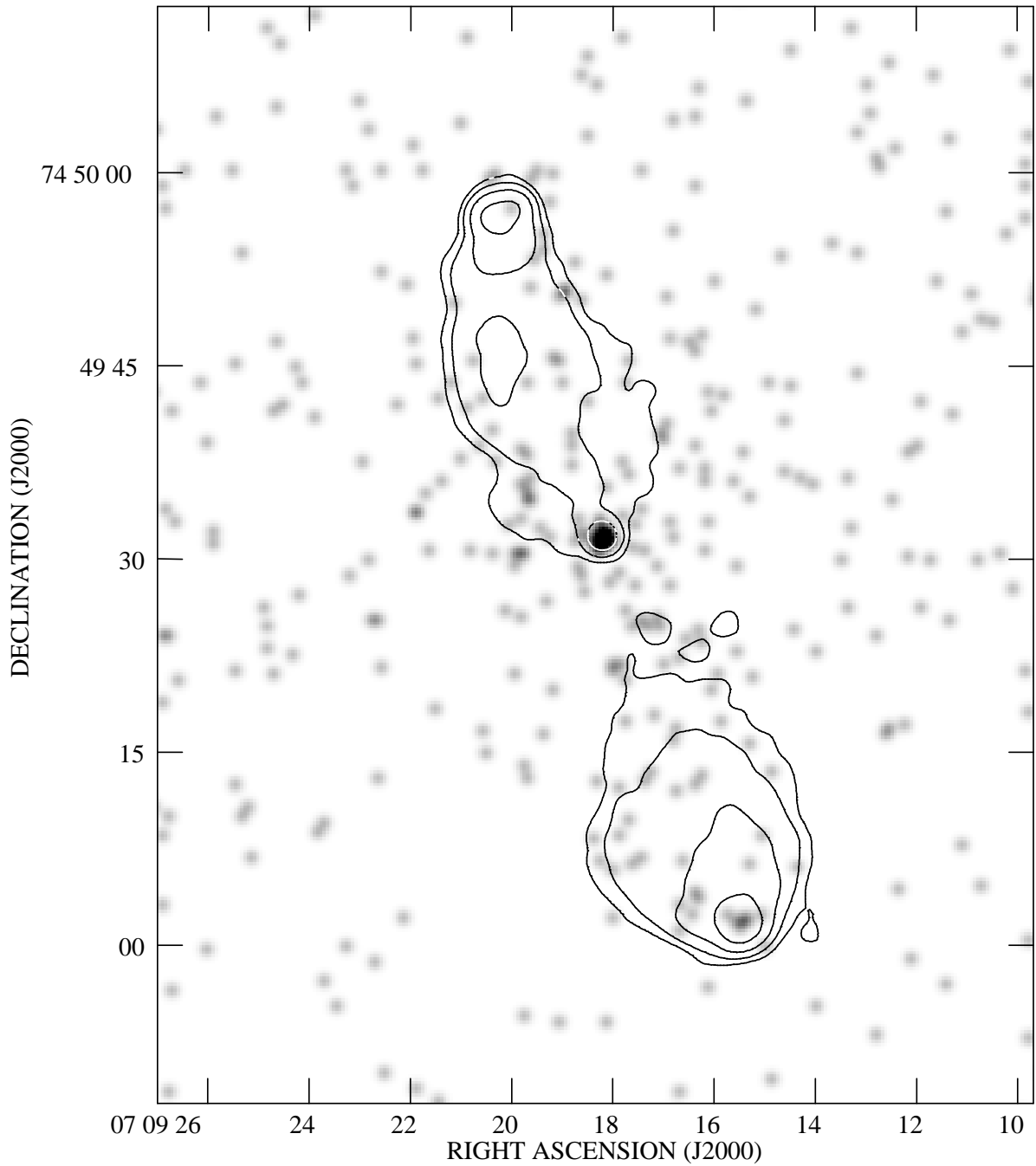}
\caption{The southern X-ray hotspot of 3C\,173.1. The greyscale shows
  the 0.5-5 keV {\it Chandra} counts smoothed with an $1\farcs0$ FWHM
  Gaussian; black corresponds to 4 counts per $0\farcs246$ pixel. The
  contours are of a $1\farcs7$ resolution 8.4-GHz VLA map taken from
  \citet{hapr97}, and are at $0.2 \times (1,4,16\dots)$ mJy
  beam$^{-1}$.}
\label{3c173.1}
\end{center}
\end{figure}

There is a weak detection of the S hotspot of this low-excitation
radio galaxy (Fig.\ \ref{3c173.1}), at well over $3\sigma$
significance. The extended emission here is again probably a
combination of inverse-Compton emission and a thermal environment. The
hotspot here is on the counterjet side.
\subsection{3C\,321}
\begin{figure}
\begin{center}
\plotone{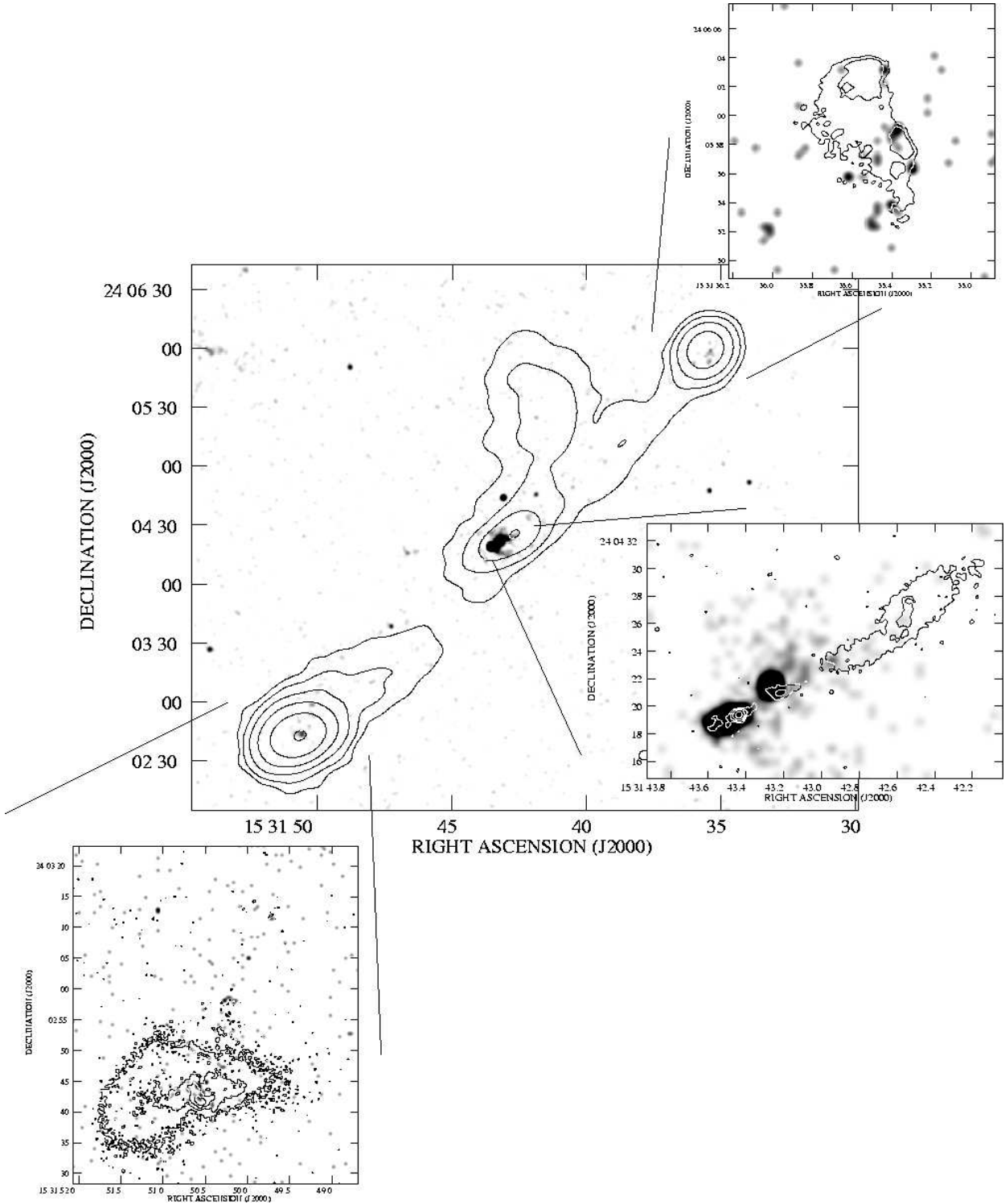}
\caption{The X-ray hotspots of 3C\,321. The main greyscale shows
  the 0.5-5 keV {\it Chandra} counts smoothed with an $2\farcs0$ FWHM
  Gaussian; black corresponds to 1 count per $0\farcs492$ pixel. The
  contours are of a 1.4-GHz VLA map with $15'' \times 13''$ resolution, and are at $2 \times (1,4,16\dots)$ mJy
  beam$^{-1}$. Insets show the same map smoothed with an $0\farcs5$
  FWHM Gaussian, and contours from a 4.8-GHz VLA map with
  $0\farcs45\times 0\farcs40$ resolution, at $0.15 \times (1,4,16\dots)$ mJy
  beam$^{-1}$; black is 1 count per pixel for the hotspots and 5
  counts per pixel for the nuclear inset.}
\label{3c321}
\end{center}
\end{figure}

Both hotspots of this nearby narrow-line FRII source are detected,
weakly but convincingly (Fig.\ \ref{3c321}). In addition, there is
X-ray emission from a weak radio jet entering the S hotspot from
slightly W of N (just visible on our high-resolution contour map). The
nuclear region shows very unusual structure. The component associated
with the radio core is extended, and, if the relative astrometry of
the {\it Chandra} and radio data is correct, shows quite strong X-ray
emission from the radio-weak jet pointing SE; the NW compact bright
X-ray source is positionally coincident not with the bright NW radio
jet, which is not clearly detected in X-rays, but with the nearby
companion galaxy seen with {\it HST} \citep{mbsw99} if we align the
center of the host galaxy with the radio core and the brightest
component in the X-ray (the {\it HST} data has the usual arcsec-scale
astrometric offset, and there are no obvious independent features with
which to align the two datasets). There is also extended emission
around the two galaxies that appears to be spatially coincident with
the known optical line-emitting material \citep{bhbv88}. The strongly
different jet-counterjet asymmetry in the radio and X-ray is hard to
explain in a model in which the jet and counterjet are intrinsically
symmetrical.

\subsection{3C\,324}
\begin{figure}
\begin{center}
\plotone{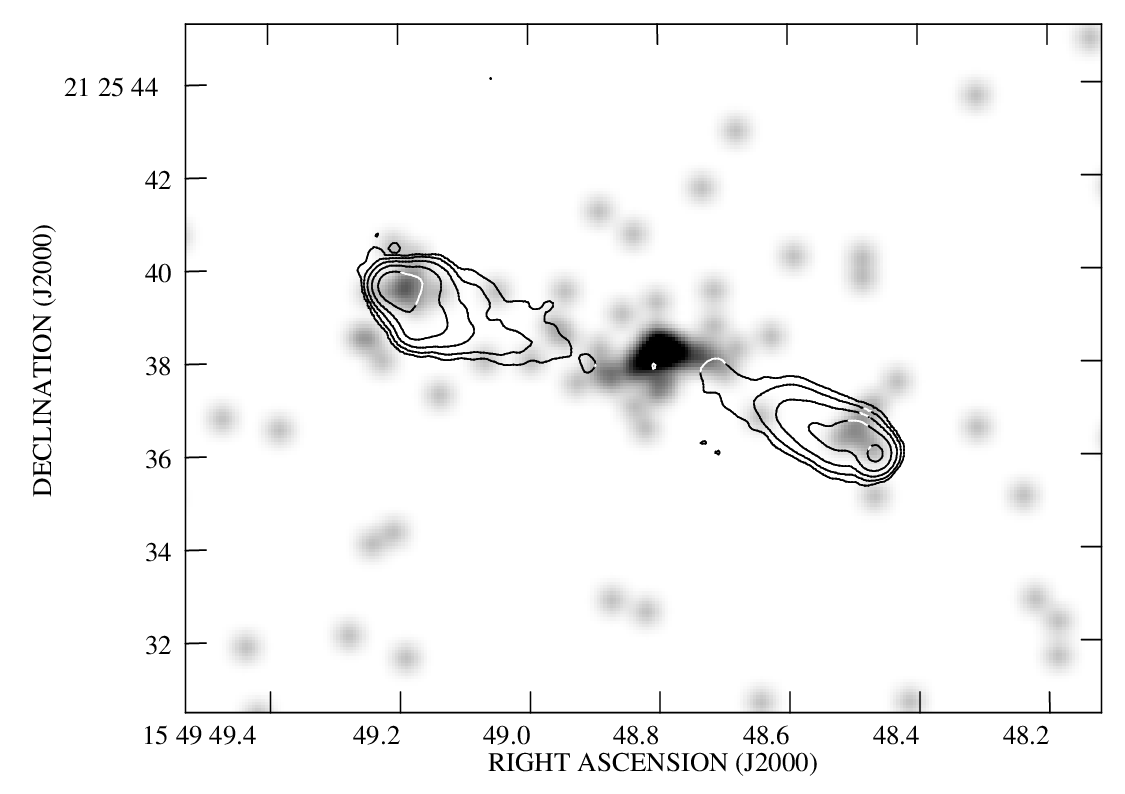}
\caption{The X-ray hotspots of 3C\,324. The greyscale shows
  the 0.5-5 keV {\it Chandra} counts smoothed with an $0\farcs5$ FWHM
  Gaussian; black corresponds to 4 counts per $0\farcs246$ pixel. The
  contours are of the 4.8-GHz VLA A-configuration map at
  $0\farcs39$ resolution, and are at $0.15 \times
  (1,4,16\dots)$ mJy beam$^{-1}$.}
\label{3c324}
\end{center}
\end{figure}

Both hotspots of this small narrow-line radio galaxy are detected
(Fig.\ \ref{3c324}), the E hotspot clearly, the W one more marginally.

\subsection{3C\,452}
\begin{figure}
\begin{center}
\plotone{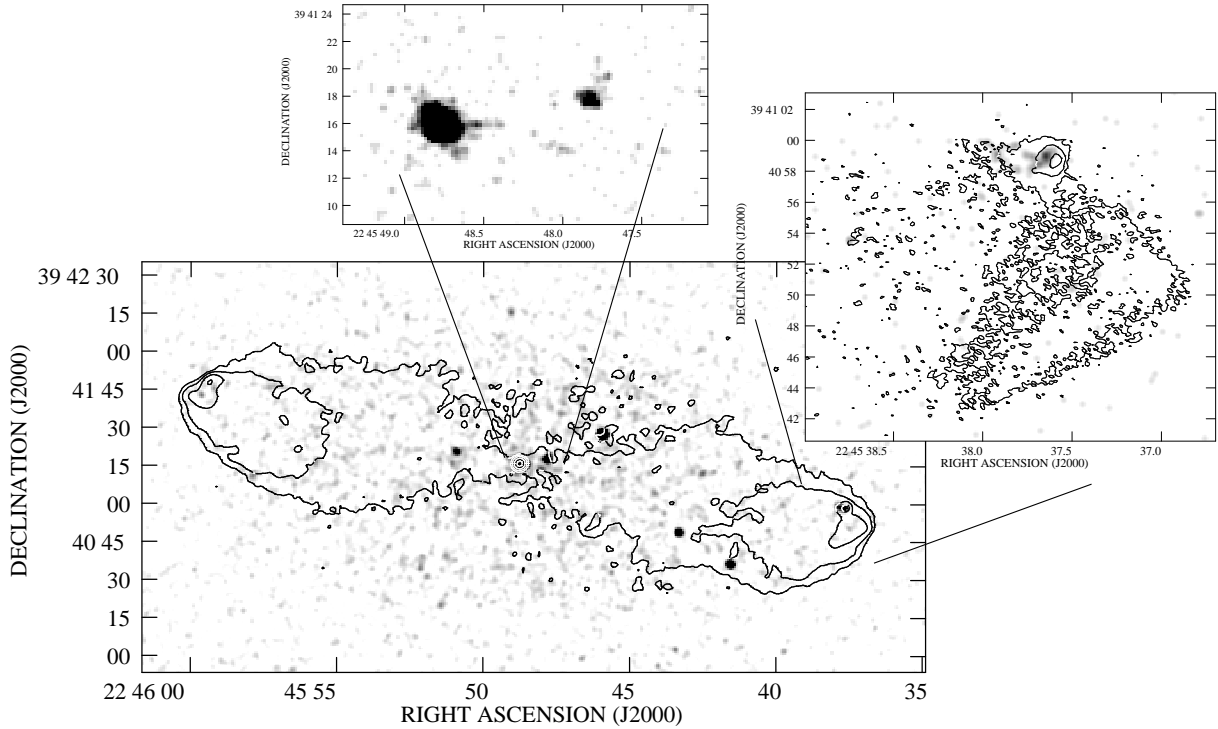}
\caption{The western X-ray hotspot of 3C\,452. The main greyscale shows
  the 0.5-5 keV {\it Chandra} counts smoothed with an $2\farcs0$ FWHM
  Gaussian; black corresponds to 10 counts per $0\farcs246$ pixel. The
  contours are of an 8.4-GHz VLA map with $2\farcs5$ resolution from
  \citet{bblp92}, and are at $0.3 \times (1,4,16\dots)$ mJy
  beam$^{-1}$. Insets show the same X-ray image smoothed with a
  $0\farcs5$ FWHM Gaussian, with black being 1 count per $0\farcs246$
  pixel. Right, the hotspot: contours are of an 8.4-GHz VLA map with $0\farcs25$ resolution from
  \citet{bblp92}, and are at $0.1 \times (1,4,16\dots)$ mJy
  beam$^{-1}$. Top, the inner jet (no contours are shown).}
\label{3c452}
\end{center}
\end{figure}

The W hotspot of this low-redshift narrow-line radio galaxy is clearly
detected (Fig.\ \ref{3c452}), as is a faint linear X-ray feature
pointing W from the nucleus towards the known radio jet in the W lobe
(though the features shown in Fig.\ \ref{3c452} have no detected radio
counterparts on high-resolution maps). Extended emission associated with the
lobes has already been reported and is modeled in terms of
inverse-Compton emission \citep{itmi02}.
\clearpage
\section{Synchrotron and inverse-Compton emission}

To guide the reader in interpreting the physics of synchrotron
and inverse-Compton emission, we include here a brief sketch of the
underlying physics. In practice these calculations are carried out by
the computer code discussed in the paper and in \cite{hbw98}, but it is useful
to set out the theoretical underpinning of the code's results and to
outline the key dependences of the model parameters.

We take as a fiducial assumption (which can then be tested by observation) the equipartition of energy between electrons (or more generally
particles of all kinds) and magnetic field. If the electron energy
spectrum (number per unit energy per unit volume) is described by a function $N(E)$, then equipartition implies
(in SI units)
\begin{equation}
{{B^2} \over 2\mu_0} = \int_{E_{\rm min}}^{E_{\rm max}}
E N(E) {\rm d}E + u_{\rm NR}
\end{equation}
where $E_{\rm min}$ and $E_{\rm max}$ give the range of electron
energies, $B$ is the magnetic field strength, $\mu_0$ is the
permeability of free space and $u_{\rm NR}$ is the energy density in
non-radiating particles. It is conventional to let $\kappa$ be the
ratio of the energy densities in non-radiating and radiating
particles: then
\begin{equation}
{{B^2} \over 2\mu_0} = (1 + \kappa) \int_{E_{\rm min}}^{E_{\rm max}}
E N(E) {\rm d}E
\label{equip}
\end{equation}
It is easy to see that the $(1+\kappa)$ term can also be used to describe an
arbitrary departure from equipartition between the electrons and
magnetic field. Our fiducial assumption is equivalent to $\kappa = 0$.

Now let us consider for simplicity a power-law
distribution of electron energies, $N(E){\rm d}E = N_0 E^{-p} {\rm
d}E$. Then the integral can be carried out analytically:
\begin{equation}
{{B^2} \over 2\mu_0} = (1 + \kappa) N_0 I
\end{equation}
where
\[
I = \left\{ \begin{array}{ll}\ln(E_{\rm max}/E_{\rm min})&p=2\\
{1\over{2-p}} \left[E^{(2-p)}_{\rm max}-E^{(2-p)}_{\rm min}\right]&p\neq
2\\
\end{array}\right .
\]
In practice, as described in the text, we may use more complicated
electron energy spectra, and then it is easiest to determine $I$
numerically.

The volume synchrotron emissivity of the ensemble of electrons
at a given source-frame frequency $\nu$ may be written \citep[e.g.,][]{l94}
\begin{equation}
J(\nu) = 
{{\sqrt{3} Be^3 \sin \theta}\over{4\pi\epsilon_0 c
m_e}} \int_{E_{min}}^{E_{max}} F(x) N(E) {\rm d}E
\label{emissi}
\end{equation}
Here $m_e$ is the mass of the electron, $e$ is its charge and $c$ is
the speed of light; $\epsilon_0$ is the permittivity of free
space. $\theta$ is the pitch angle of the electrons with respect to the
magnetic field direction and $x$ is defined by
\[
x = {{4\pi m^3c^4}\over{3e}}{{\nu}\over{E^2 B \sin \theta}}
\]
$F(x)$ is a sharply peaked function of $x$, reflecting the fact that
electrons of a given energy radiate at a well-defined
frequency;
\begin{equation}
F(x) = x \int^\infty_x K_{5/3}(z) {\rm d}z
\end{equation}
where $K_{5/3}$ is the modified Bessel function of order $5/3$.
Assuming pitch angle isotropy, we can integrate equation
\ref{emissi} over pitch angle and our assumed electron power law
\citep[e.g.,][]{l94} to find that
\begin{equation}
J(\nu)  = C N_0 \nu^{-{{(p-1)}\over 2}} B^{{(p+1)}\over 2}
\label{jnu}
\end{equation}
where
\[
C = c(p) {{e^3} \over {\epsilon_0 c m_e}} \left({{m_e^3 c^4}\over {e}}\right)^{-
(p-1)/2}
\]
and $c(p)$ is of order 0.05 and depends only weakly on $p$. Since
equation \ref{jnu} describes a power law in frequency, the electron
energy index $p$ can be determined by observation: typical values lie
in the range 2--3, and our assumption of a low-frequency spectral
index $\alpha = 0.5$ corresponds to $p=2$.

If we know
the emissivity $J$, we can use equation \ref{jnu} to eliminate $N_0$
from equation \ref{equip}
\begin{equation}
{{B^2} \over 2\mu_0} = (1+\kappa) {{J(\nu)}\over C}
\nu^{{{(p-1)}\over 2}} B^{-{(p+1)}\over 2} I
\label{bsolve}
\end{equation}
and we can now solve for $B$:
\begin{equation}
B = \left[2\mu_0 (1+\kappa) {{J(\nu)}\over C}
\nu^{{{(p-1)}\over 2}}
I\right]^{2\over p+5}
\label{bval}
\end{equation}
In practice, we take account of the fact that $J(\nu)$ is not always a
power law by numerically integrating equation \ref{emissi} and then
solving equation \ref{bsolve} numerically with a root-finding
algorithm, but the main dependences are encapsulated in equation
\ref{bval}. Since the energy density is proportional to $B^2$, we can
see that it increases as $(1 + \kappa)^{4/p+5}$; thus, a non-zero
value of $\kappa$ affects the magnetic field strength in the expected
sense. Moreover, we can now substitute back into equation \ref{jnu} to
eliminate $B$: this gives
\begin{equation}
J(\nu)^{4\over {p+5}} = C^{4\over {p+5}} N_0 \nu^{-\left({2(p-1)\over{p+5}}\right)} \left
  [2\mu_0 (1+\kappa) I\right]^{{p+1}\over{p+5}}
\end{equation}
and, since $J(\nu)$ and $\nu$ are known and constant for a given
observation and a known source geometry, we can see that the number density of
electrons is expected to decrease with increasing $\kappa$:
\begin{equation}
N_0 \propto (1+\kappa)^{-{{p+1}\over{p+5}}}
\label{n0k}
\end{equation}
If the geometry is doubtful, the calculated emissivity is a function of volume: $J
\propto S/V$, where $S$ is the observed flux density. So we expect
\begin{equation}
N_0 \propto V^{-{4\over p+5}}
\label{n0v}
\end{equation}

For sphere of radius $r$ with uniform particle and magnetic field density, the synchrotron self-Compton emissivity at a given frequency
$\nu_1$ is given by \citep{hbw98}
\begin{equation}
J_{\rm ic}(\nu_1) = {9\over 16} m_{\rm e}^2 c^4\nu_1\sigma_{\rm T} r\int^{E_{\rm
 max}}_{E_{\rm min}} \int^{\nu_{\rm max}}_{\nu_{\rm min}}{{N(E) J(\nu_0)}\over{E^2
\nu^2_0}} f(x) {\rm d}\nu_0 {\rm d}E 
\end{equation}
where $m_e$ is the electron mass, $c$ is the speed of light,
$\sigma_T$ is the Thomson cross-section, $J(\nu_0)$ is the synchrotron
emissivity as a function of frequency, and $f(x)$ is a function of
$E$, $\nu_1$ and $\nu_0$ defined by \cite{rl79}. The
code we use performs this integration numerically for a given
synchrotron and electron energy spectrum, and also carries out the
similar calculation for illumination from the microwave background
radiation; an analytical form of the integral for power-law electron
and photon distributions could be derived but is not necessary here.
The key feature of this equation is that the inverse-Compton
emissivity (which determines the predicted IC flux density and thus
$R$ for a given observed flux) is linear in the number density of
electrons $N(E)$, and thus linear in $N_0$ for the power-law analysis
we have described above. For a given source, with known spatial and
spectral properties, it is the dependence of $N_0$ on $(1+\kappa)$
given by equation \ref{n0k} that primarily determines the value of
$R$. (If the spectrum is not a pure power law, the change in the form
of $J(\nu_0)$ as a result of the change in $B$ also has a
non-negligible effect.)

If the volume is not known, for example
because of a low filling factor, then the dependence of $J_{\rm ic}$
on volume can be determined from above:
\begin{equation}
J_{\rm ic} \propto V^{{1\over3}} V^{-{4\over p+5}} V^{-1} =
V^{-{2\over 3} - {4\over p+5}}
\end{equation}
and this means that the observed inverse-Compton flux density, which
is proportional to $V J_{\rm ic}$, goes as $V^{1/3 - 4/(p+5)}$. This is
a very weak dependence for plausible $p$ values: for $p=2$, $S_{\rm
ic} \propto V^{-5/21}$.

\end{document}